\documentclass[12pt]{article}

\usepackage{amssymb}

\usepackage[utf8]{inputenc}
\usepackage[T1]{fontenc}
\usepackage{url}
\usepackage{booktabs}
\usepackage{amsfonts}
\usepackage{nicefrac}
\usepackage{microtype}
\usepackage{xcolor}

\usepackage{amsmath}
\usepackage{graphicx}

\makeatletter
\let\MyIntOrig\int
\def\MyIntSpace{\hspace{-.3em}}
\def\int{\MyInt}
\def\MyInt{\MyIntOrig\MyIntSkipMaybe}
\def\MyIntSkipMaybe{
  \@ifnextchar_{\MyIntSkipScript}{%
  \@ifnextchar^{\MyIntSkipScript}{%
  \@ifnextchar\limits{\MyIntSkipTok}{%
  \@ifnextchar\nolimits{\MyIntSkipTok}{%
  \MyIntSpace}}}}%
}
\def\MyIntSkipScript#1#2{#1{#2}\MyIntSkipMaybe}
\def\MyIntSkipTok#1{#1\MyIntSkipMaybe}
\makeatother

\newcommand{\Expect}{{\rm I\kern-.2em E}}

\usepackage{mathtools}

\usepackage{amsthm}

\usepackage{verbatim}

\newcommand{\minus}{\scalebox{0.8}{-}}
\newcommand{\plus}{\scalebox{0.6}{\!+}}

\usepackage{bbm}
\usepackage{nicematrix}
\usepackage{xcolor}

\newtheorem{prop}{Proposition}
\newtheorem{defin}{Definition}

\newtheorem{examp}{Example}

\newtheorem{assumption}{Assumptions}
\newtheorem*{assumption*}{Assumptions}

\newtheorem{princ}{Principle}

\usepackage{stackengine}
\newcommand\stackequal[2]{%
  \mathrel{\stackunder[2pt]{\stackon[4pt]{$\to$}{$\scriptscriptstyle#1$}}{%
  $\scriptscriptstyle#2$}}}

\usepackage{nameref}

\usepackage{natbib}
\makeatletter
\def\@date{} 
\makeatother

\title{Sample Observed Effects: \\Enumeration, Randomization and Generalization}
\author{
  Andre F. Ribeiro\thanks{\texttt{ribeiro@alum.mit.edu}, Department of Applied Mathematics and Statistics, University of Sao Paulo, Sao Carlos, SP, 13560-970,  Brazil.} 
}
\begin{document}

\maketitle

\begin{abstract}
The widely used 'Counterfactual' definition of Causal Effects was derived for unbiasedness and accuracy - and not generalizability. We propose a Combinatorial definition for the External Validity (EV) of intervention effects. We first define the concept of an effect observation 'background'. We then formulate conditions for effect generalization based on samples' sets of (observed and unobserved) backgrounds. This reveals two limits for effect generalization: (1) when effects of a variable are observed under all their enumerable backgrounds, or, (2) when backgrounds have become sufficiently randomized. We use the resulting combinatorial framework to re-examine several issues in the original counterfactual formulation: out-of-sample validity, concurrent estimation of multiple effects, bias-variance tradeoffs, statistical power, and connections to current predictive and explaining techniques.

Methodologically, the definitions also allow us to replace the parametric estimation problems that followed the counterfactual definition by combinatorial enumeration and randomization problems in non-experimental samples. We use the resulting non-parametric framework to demonstrate (External Validity, Unconfoundness and Precision) tradeoffs in the performance of popular supervised, explaining, and causal-effect estimators. We also illustrate how the approach allows for the use of supervised and explaining methods in non-i.i.d. samples. The COVID19 pandemic highlighted the need for learning solutions to provide predictions in severally incomplete samples. We demonstrate applications in this pressing problem.
\end{abstract}

\section{Introduction}



Non-parametric effect estimators (such as permutation tests ~\citep{Lee:1990aa} and feature attribution methods ~\citep{10.1613/jair.1.12228}) can often only be used under strict assumptions of no unobserved factors. Consider, for example, a sample that contains units that both have received a treatment ('treatment' units) and others that did not ('control'). To estimate effects, a simple strategy is to permute treatment assignment among treatment-control pairs that are similar enough across relevant variables ~\citep{Morgan:2007aa,Rubin:2005aa}. This strategy relies, however, on the assumption that paired units are exchangeable besides their treatment status. The assumption is easily violated if units differ in any unobserved way (even in high-dimensional samples it is hard for researchers to make the claim that there are no overlooked, relevant distinctions). The main challenge is that of unobserved confounding, and of sampling conditions where uncontrolled variation of out-of-sample factors bias in-sample estimates.  Under these conditions, sample observation of effects, for example, reflect not only the true effect of sample factors, but also of uncontrolled and unobserved factors. In Experimental Design, randomization plays a crucial role in alleviating the effects of unobserved confounding ~\citep{Zhang:2023to,10.1093/biomet/56.2.231}. Randomized Experiments provide a solid foundation, and are still the golden rule, for making causal inferences about the effects of treatments in the Sciences. Non-experimental samples can be seen as experimental samples where randomization did not occur perfectly. Despite these known shortcomings, permutation and Shapley-value based feature attribution methods ~\citep{10.1613/jair.1.12228, ribeiro-growth} continue to be used in practice in non-experimental and partially observed data, with little clarity over the resulting effect estimates' generalizability and biases. We provide a combinatorial definition for effect generalizability and demonstrate how to select subsamples in non-experimental data that lead to effect estimates with high generalizability and low confoundness.  We demonstrate uses of these concepts when estimating causal effects non-parametrically or when explaining the performance of supervised predictors. The latter is currently an intensely studied subject in Machine Learning and Artificial Intelligence ~\citep{10.5555/3295222.3295230,10.1613/jair.1.12228}.

A sample's Data Generating Process (DGP) refers to the underlying mechanism or model that generated the sample. Partition the set $Z$ of DGP variables into two subsets, $Z = X_m \cup U_q$, the first with its $m$ observed variables $X_m =  \{-1,+1\}^m$, and the second with $q$ unobserved $U_q =  \{-1,+1\}^q$. Imagine also variables in $X$ and $U$ are constrained by precedence relationships, as described by two partially ordered set (posets) $G(Z)=\{G(X),G(U)\}$. The distinct ways in which $\{G(X),G(U)\}$ can be combined is described by the distinct linear extensions among these posets. For poset chains $\{G(X),G(U)\} =\{\;  \mathbf{a} < \mathbf{b}, \;\;u_1 < u_2 < u_3 \;\}$, for example, the extensions are

\begin{examp}[$U{-}X$ Linear Extensions]\label{ex-extensions}
\footnotesize$\mathbf{ab}u_1u_2u_3 \,\,\mathbf{a}u_1\mathbf{b}u_2u_3,\allowbreak \,\mathbf{a}u_1u_2\mathbf{b}u_3, \allowbreak \,\mathbf{a}u_1u_2u_3\mathbf{b}, \allowbreak u_1\mathbf{ab}u_2u_3 \, \,u_1\mathbf{a}u_2\mathbf{b}u_3, \,u_1\mathbf{a}u_2u_3\mathbf{b},\, u_1u_2\mathbf{ab}u_3 \,\, u_1u_2\mathbf{a}u_3\mathbf{b},\, \allowbreak u_1u_2u_3\mathbf{ab},\,\allowbreak \mathbf{a}u_1u_2u_3\mathbf{b}.$
\end{examp}

 The different extensions can be seen as different confounding conditions under which sample factors $X$ can be observed.  We imagine real-world samples can be collected under any such conditions, and we would like to formulate statistical properties for effects across partially observed samples, $q>0$. The likelihood of a particular $U{-}X$ extension follow, in turn, relationships between the DGP probability distributions for $X$ and $U$,

\begin{examp}[$U{-}X$ Data-Generating Order Relations]\label{ex-dgp}
\footnotesize\begin{align*}	
&\underset{\vphantom{\text{('fast')}}\smash{\text{('fast')}}}{a, b < u_i\rule[-1.5ex]{0pt}{0pt}} \quad : \;\frac{p_t(b\,|\,a)}{p_t(u_i\,|\,a)}\gg 1, \quad\; \underset{\vphantom{\text{('in-sync')}}\smash{\text{('in-sync')}}}{a, b \lessgtr u_i\rule[-1.5ex]{0pt}{0pt}} :\; \frac{p_t(b\,|\,a)}{p_t(u_i\,|\,a)}\approx 1, \quad\; \underset{\vphantom{\text{('slow')}}\smash{\text{('slow')}}}{a, b > u_i\rule[-1.5ex]{0pt}{0pt}}: \;\frac{p_t(a\,|\,b)}{p_t(u_i\,|\,a)}\ll 1, 
\end{align*}
\end{examp}

 
where $p_t(b\,|\,a)$ and $p_t(u_i\,|\,a)$ denote the conditional probability of factors $b \in X$ and $u_i \subseteq U$, given the observation of factor $a$ at a time $t=0$, and $\lessgtr$ denotes poset incomparability (when either $a < u_i$ or $a > u_i$ can hold true). The first condition implies that the partial order $a < b  < Z-\{a\}$ will apply with high likelihood in the resulting sample. The condition corresponds to the first extension in \textit{Example (\ref{ex-extensions})}. That is, that there are no confounders between subsequent $a$ events. While simplifying estimation in this case, the same only holds true for specific subsamples in the other two scenarios (in equal proportion across all factors in the second and in rare subsamples in the third). For short, we refer to $U{-}X$ systems as being under \emph{fast}, \emph{in-sync} and \emph{slow} sampling in these respective conditions. We define sampling strategies for cause-confounder separability (i.e., strategies that generate samples where causes and confounders can be distinguished) across each such condition.  


 


\subsubsection{Alternative Orders and Confounding: A Preliminary illustration}\label{sect-obscf}

For illustration, consider the effects observed in a sample with a true cause $a$, a confounder $b$ (with effects on $y$ fully correlated with its root cause $a$), and a spurious variable $c$, $X=\{a,b,c\}$. We will use Latin square diagrams ('squares') as aid to illustrate several concepts in this article. Fig.\ref{fig-intro}(a) shows a $4$-factor square. Each square is associated with a sample unit $x_0$, its \emph{reference} unit, which stratifies the sample (i.e., other units) across its cells. A square with reference $x_0$ correspond to the set of effect observations with singleton factor differences,

\begin{equation*}
\begin{split}
 \bigcup_{x \in \mathcal{P}(X)} {{\Delta}} y(x - x_0 = \{a\}), \qquad \forall \,a \in X.  
\end{split}
\end{equation*}

where $\mathcal{P}(X)$ is the powerset (set of sets) of $X$ and ${\Delta} y(x - x_0)$ is an observation of effect in an outcome-of-interest, $y \in \mathbb{R}$, after an intervention on factors $x - x_0$. A complete square contains therefore $\binom{m}{2}=\frac{m \cdot (m-1)}{2}$ effect observations, and a sample (or all its squares) is limited to $2^m$ observations. We will say that squares' diagonals enumerate, for example, observations of effect for fixed factors across different (observed) 'backgrounds'. For the cause $a$ and confounder $b$, differences in effect observations across diagonals are

\begin{examp}[Iterative higher-order effects]\label{ex-diffs}
\def\arraystretch{0.4}\scriptsize	
\[
\begin{array}{lll @{\hskip 30pt} lll}
\multicolumn{3}{c}{\textrm{diagonal}\; a}&\multicolumn{3}{c}{\textrm{diagonal}\; b }\\[8pt] 
 \Delta {y}( a)&&&  \Delta {y}( b)&&\\
      & \Delta {y}( a\, |\, b){-}\Delta {y}( a ){=}0&&& \Delta {y}( b \,| \,a){-}\Delta {y}( b ){=}\Delta {y}( a)& \\
\Delta {y}( a \,|\, b) &             &{=}0, &\Delta {y}( b \,|\, a)&&{=}{-}\Delta {y}( a),\\
    & \Delta {y}( a  \,|\, bc ){-}\Delta {y}( a | b ){=}0&&&\Delta {y}( b\, |\, ac ){-}\Delta {y}( b \,|\, a ){=}0&  \\
 \Delta {y}( a\, |\, bc ) &&& \Delta {y}( b \,|\, ac ) &&      \\
\end{array}
\]
\def\arraystretch{1}
\end{examp}

We see that when the cause $a$ is in the sample, it sustains its effects throughout arbitrary sample imputations (indicated by the conditional statements), while the confounder also assumes negative values. In a square with factors $X$, observed under different orders, effect observations for confounders change many times, while for causes they remain invariant throughout the observed range of background variation. 

More importantly, however, this indicates that, to identify confounders we need samples where we, minimally, observe the effect of both adding factor $a$ to units with $b$, and of adding factor $b$ to units with $a$ (i.e., under alternative orders). These conditions are necessary because $b$ has no effects only in the latter case, $ \Delta y (a) + \Delta y (a \,|\, b) >  \Delta y (b) + \Delta y (b \,|\, a)$. Samples that include these conditions allows us to see, from effect observations alone, that $b$'s effects were brought by $b$’s correlation with $a$, and not its own independent effect on $y$. 


\begin{princ}[Cause-Confounder Separability] \label{prop-sep}
For any factors $a,b \in Z$, observing their effects under conditions $\Delta y(a\,|\,b)$ and $\Delta y(b\,|\,a)$, or under ordinal relations $a>b$ and $b>a$, is necessary to identify, from effect observations, whether $a$ confounds $b$ (and vice versa). When observed under such conditions, effect variances, $\textrm{Var}[\Delta \hat{y}(b)]$, for causes, confounders and spurious factors become distinct, and have expected values of $0$, $\nicefrac{\rho_{ab}}{4} \Delta y(a)$, $\nicefrac{1}{2}\Delta y(a)$, where $\rho_{ab}$ is the Spearman's rank correlation coefficient between confounder and its root cause, and $\Delta y(a)$ the root cause's effect.  
\end{princ}

According to the principle, under specific ordinal conditions on $U{-}X$ factors, effect observations for confounders have an expected effect variance of $0.25\times \rho_{ab}$ (i.e., the number of times the confounder appears before its root cause in square rows) of the true effect $\Delta y(a)$ and causal variables $0$. One such condition is of a completely observed sample, $q=0$. In this case, the principle is associated with the observation of each effect under the cyclic permutation of all others factors (as illustrated by distinct square rows). The principle states then that effect variances, $\Expect[ \,\textrm{Var}[\Delta \hat{y}(a)] \,]$, are a sufficient statistic to identify confounders under a fixed level of cause-confounder correlation $\rho_{ab}$. Conditions when $q>0$ are established below.



\subsection{Effect estimates and Unobserved Sample Variation}\label{sect-eff}


Before formulating necessary conditions for the generalizability and confoundness of effect observation in generic samples, we formulate our working definition of effects, their generalizability, and a roadmap to the article. According to the Potential Outcomes framework definitions for \emph{what is} a causal effect ~\citep{Morgan:2007aa,Rubin:2005aa}, if $y$ is an outcome-of-interest and $a$ is a treatment indicator, then the causal effect of $a$ is the difference

\begin{equation}\label{eq-rubin}
\Delta y(a) = y_i^{\texttt{+}a} - y_i^{\texttt{-}a},
\end{equation}

where $y_i^{\texttt{+}a}$ is the outcome of individual $i$ under the treatment, and, $y_i^{\texttt{-}a}$ without the treatment, $a \in X$. The central concept behind Eq. (\ref{eq-rubin}) was inspired by experimental estimation: by fixing every factor, other than the treatment, we can declare that the observed difference in outcome was certainly caused by the treatment, and the treatment alone. The definition is an ideal, as it is impossible to observe outcomes for an individual, concurrently, in two different and totally fixed conditions. The theory goes that we may, instead, 'fix' factors in expectation, and across individuals. If the treated and non-treated subpopulations have the same expected values across all factors then any difference between the groups is due to the treatment, given large enough samples. This can be paraphrased with an independence statement: \textbf{treatment assignment} must be  independent on all outcome-relevant factors. This rationale led to the notion of a sample's 'balance' in non-experimental estimation ~\citep{Rubin:2005aa,Morgan:2007aa} and the objective most current counterfactual causal effect estimators maximize. 

 While not a lot is known theoretically about the Counterfactual definition ~\citep{Abadie:2006aa,pmlr-v70-shalit17a}, we can observe that for Eq.(\ref{eq-rubin}) and a pair of sample units $x_i$ and $x_j$, 

\begin{equation}\label{eq-acc}
\textrm{Var}(x_i{-}x_j) = \textrm{Var}(x_i)+\textrm{Var}(x_j) - 2\textrm{Cov}(x_i,x_j).
\end{equation}

A maximally accurate estimator (i.e., one with minimum variance) minimizes the covariance between sample units it is comparing.  There is, however, a problem with this definition. It is only guaranteed to hold under a very strict set of conditions. Namely, it holds only if all factors of relevance are held constant among individuals. Such estimates are poised to not generalize across populations.

We consider an effect, instead, as 

\begin{align}\label{eq-eff-def}
\Delta y(a)& = \frac{1}{m}\sum\limits_{\pi \in \Pi_n(Z/\{a\})} \Big[ \,y_{>}(a\,|\,\pi) - y_{\leq}(a\,|\,\pi)\, \Big],
\end{align}


where $y_{\leq}$ is the observed outcome $y$ of a sample unit with, exclusively, the set of factors before $a$ in the permutation order $\pi$, and $y_{>}$ of units also containing the factors after $a$. Difference between units in Eq.(\ref{eq-eff-def}) constitutes observations of effects of the same factor in distinct \emph{backgrounds}, $\pi \in \Pi_n(Z/\{a\})$. Each background fixes a set of factors for their 'treatment' (before $a$ in $\pi$, or, sample effects that do not reflect the effect of $a$) and 'control' (after $a$) for sample units.  

Furthermore, $\Pi_n(Z/\{a\})$ denotes a subset of the $(m+q)!$ full set of permutations of factors $Z$ possible between sequential observations of $a$. Which subsets can offer guarantees over the generalizability of sample effect estimates? We formulate the concept of External Validity increasing (EV-increasing) permutations as those that can increase the generalizability of effects in samples. The article's main argument is structured as follows, 

\begin{center}
\begin{tabular}{|p{0.25cm} p{4cm}@{\hspace{5mm}} | @{\hspace{5mm}} p{8cm}|}\hline
 & \textbf{Section} & \textbf{Summary} \\
\hline
\ref{sect-gaps}. & \textit{\nameref{sect-gaps}}  &  formulates the concept of sample 'gaps' to describe the action of confounders $U$ across in-sample factors $X$. \\
\hline
\ref{sect-assign}. &  \textit{\nameref{sect-assign}} & formulates permutation subsets $\Pi_n(Z/\{a\})$ that are EV-increasing and CF non-increasing, and their relationship to sample gap statistics. \\
\hline
\ref{sect-backrandom}. & \textit{\nameref{sect-backrandom}} & formulates non-parametric tests for effect background randomization based on the previous concepts. \\
\hline
\ref{sect-related}. &  \textit{\nameref{sect-related}}  & recasts the approach under graphical and structural causal frameworks and relate it to other non-parametric effect estimators. \\
\hline
\ref{sect-power}. &  \textit{\nameref{sect-power}}  & formulates resulting sample size requirements for effect generalizability. \\
\hline
\end{tabular}
\end{center}




Based on these concepts, we say an effect is \textbf{Externally Valid (EV)} if it holds across the subset $\Pi_{n}(Z/\{a\})$ of backgrounds,

\begin{equation}\label{eq-ev-def}
\textrm{EV}(a) = \textrm{Var}^{-1}\Big[\Delta y(a) \,|\, \Pi_{n}(Z/\{a\}) \Big].
\end{equation}


EV is here the inverse of the variance (i.e., the precision) of observed effects under all (e.g., asymptotically) EV-increasing backgrounds, $\Pi_{n}(Z/\{a\})$, for a treatment $a$. This \textbf{defines effects in a way that is almost opposite to Eq. (\ref{eq-rubin})}. It calls for effects to be observed under large variation, as opposed to no variation.  According to Eq.(\ref{eq-eff-def},\ref{eq-ev-def}), the counterfactual ideal in Eq.(\ref{eq-rubin}) is, in fact, the estimate with \textbf{minimal EV}. Underlying these definitions and \textit{Principle (\ref{prop-sep})} is the observation that, unlike confounders, causes can be fundamentally defined as 'perfect controls' ~\citep{RefWorks:doc:5911ec76e4b0ac17f7d9f458}. Similar definitions have been explored in model robustness research ~\citep{Peters:2016aa,Buehlmann:2020aa}. The non-parametric definitions here frame the problem, in particular, not as simply the invariance (of effects), but invariance of effects under  maximum variance of their 'backgrounds'. Shifting the focus to the later, $\Pi_n(Z/\{a\})$, is a key but challenging part of the proposed solution. 


\begin{figure}
\centering
\includegraphics[width=1\linewidth]{figs/final/fig-1-6.pdf}\\
\caption{(a) $4{\times}4$ Latin-Square ('square') as sets of effect observations for a sample unit $x_0 \subseteq X$, (b) $3$ samples (horizontal lines, top) with increasing dimension $m = \{1,2,4\}$ and their ordinal statistics (interval ticks) , as well as alternative samples of confounding unobserved factors $U$ (circles) (top), resulting distribution and percentage of enumerable permutations of effect observations across ordinal variables (bottom and right).}\label{fig-intro}
\end{figure}

\section{Ordinal Statistics and Sample Confounding Gaps}\label{sect-gaps}


If $\{X_a, X_b, \ldots, X_{m}\}$ are random samples from a particular DGP distribution of $X_m$, it will be important to know whether occurrences of a factor $a$ are expected to follow or precede occurrences of another factor $b$. The likelihood of such order relations is determined by the DGP, as illustrated in \textit{Example (\ref{ex-dgp})}. To describe such patterns more formally, we define $X_{(1)}, X_{(2)}, \ldots, X_{(m)}$ to be an ordered version of sample factors ~\citep{10.5555/1717330}, that is,

\begin{align*}
X_{\min} \equiv X_{(1)} < X_{(2)} <\ldots < X_{(m)} \equiv X_{\max} 
\end{align*}

Notice that $X_{k}$ are random variables, $0<k\leq m$. For simplicity, we assume that, in sufficiently large samples and time resolution, ranks do not coincide, leading to strict inequalities (as opposed to $\leq$). 

Suppose that $X_1, X_2, \ldots, X_{m}$ follow a continuous distribution with PDF $P$. We will use $X$, $U$ or $Z$ as alternative labels when considering distributions over observed, unobserved or either factors.  Imagine then a random sample of $U$ factors whose minimum is below an in-sample value $x$. Fig.\ref{fig-intro}(b) give examples where circles are $U$ samples and interval ticks are $X$ samples. The successive lines show $X_{(k)}$ in samples with increasing dimensionality $m$. Circles to the left of the dashed line, and $X_{(1)}$, introduce irreducible unobserved confounding effects to the sample. Under i.i.d. sampling, this can be expressed as

\begin{align}\label{eq-min}
P(x \, \text{is 'confounded' by } U) &= P(\text{at least one factor in } U \text{ is} \leq x)\\ &= P(U_{(1)} \leq x)  \nonumber
\end{align}.



We are specially interested in the distribution of the $k$-th sample \emph{gap} and its associated variables,

\begin{align}\label{eq-gaps}
\Delta_{(k)}X \stackrel{\text{def}}{=} X_{(k)}-X_{(k+1)}&,\qquad \Delta_{(k)}y(X{=}x) \stackrel{\text{def}}{=} Y_{(k)}(X{=}x) -Y_{(k+1)}(X{=}x),\nonumber\\ 
\Delta U_{(k)} \stackrel{\text{def}}{=} &P(u_{(k)} \,|\, X_{(k)} < u < X_{(k+1)} ),
\end{align}

where $u \in \mathcal{P}(U)$. In this notation, $\Delta X_{(k)}$ and $\Delta y_{(k)}$ are gap and gap-effect distributions, and $\Delta U_{(k)}$ is the distribution of unobserved confounders across gaps. We also define the counts 

\begin{align*}
 \Delta_{(k)}m \stackrel{\text{def}}{=}  | \,a \in \text{{Range}}(X_{(k)}) \mid P(X_{(k)}{=}a) > 0 \, |, \\ \Delta_{(k)}q \stackrel{\text{def}}{=}  | \,u \in \text{{Range}}(\Delta U_{(k)}) \mid P(U_{(k)} = u) > 0 \, |,
\end{align*}

where $\Delta_{(k)}m$ and $\Delta_{(k)}q$ are the number of, respectively, observed and unobserved factors with positive probability in the sample's $k$-th ordinal position or gap. These gap statistics are closely related to the possible extensions among posets $\{G(X),G(Y)\}$ (such as those illustrated in \textit{Example (\ref{ex-extensions})}).


Sample gap effect statistics will become relevant because they \textbf{reflect only the effect of new causes introduced within the gap} ($\Delta y(u)$, $a<u_k< b$ for two sample factors $a$ and $b$ and unobserved causes $u_k \subseteq U$). Of particular interest are gap effects when in-sample factors are the same in subsequent effect observations.  In this case, outcome differences in a gap $k$ reflect exclusively the effect of unobserved factors 'falling' in $k$, $0 < k \leq m$. In a square diagram with $X=\{a,b,c\}$, for example, $Y_{(k)}(X{=}a)-Y_{(k+1)}(X{=}a)$ denotes differences in outcome across its main diagonal, Fig.\ref{fig-intro}(a). Eq.(\ref{eq-min}) thus suggests distinct levels of confounding across a sample's effect observations,

\begin{prop}[Sample Ordinal Statistics and Effect Variance]\label{prop-varordinal}
Effect observations of factor $a$ at $\textrm{argmin}_k X_{(k)}=a$ are a sample's minimally confounded and maximally precise individual observations of $a$'s effect, $\textrm{Var}^{-1}[\Delta y(a)]$. In a single-row square, the effect of its first factor is observed at maximum precision. All other effects can be estimated confounded only by $\Delta_{(k)}y$, by conditioning on effect observations prior in the square row order (and their interaction effects). 
\end{prop}

\begin{prop}[Full-Observability and Effect Variance]\label{prop-varordinal2}
 In a square with $m$ factors and $q=0$, effects of all factors $X_m$ can be observed, simultaneously, at maximum precision and unconditionally.    
\end{prop}

The first proposition suggests that, even when a factor $b$ is not observed at a minimal order $k=1$, unobserved confounding effects at $k=1$ can still be discounted from estimates by using other sample factor effects prior to $b$ in row order. For example, when $a$ is observed at $k=1$ but $b$ only at $k=2$, $ \textrm{Var}[\Delta y_{(2)}(b\,|\,a)-\Delta y_{(1)}(a)] \leq  \textrm{Var}[\Delta y_{(2)}(b\,|\,a)]$.


We will consider, in particular, the relationship between the number of permutations that $U{-}X$ systems can generate and their impact in the statistics of effects across sample gaps. According to \textit{Principle (\ref{prop-sep})} the first is a requirement for the separability of causes and confounders, when using effect observations.  Some distributions of unobserved causes across gaps can asymptotically generate all distinct orders (all permutations of unobserved factors) and some can't. This can be described by a multinomial coefficient, or, by dividing the set of possible permutations in a $U{-}X$ system, $\Pi_n(Z/\{a\})$, and the ideal full set of $q!$ permutations of $U$,

\begin{align}\label{eq-gapprob}
\frac{|\Pi_n(Z/\{a\})|}{q!}  &\stackequal{}{t\rightarrow \infty}  
  \frac{\prod_{i=1}^l{\Delta q_{(i)}!}}{\left(\sum_{i=1}^l{\Delta q_{(i)}}\right)!}
  = \frac{\Delta q_{(1)}!\, \Delta q_{(2)}!\, \cdots\, \Delta q_{(m)}!}{q!} \nonumber \\&= {q \choose \Delta q_{(1)}, \Delta q_{(2)}, \ldots, \Delta q_{(m)}}^{-1}.
\end{align}


This can be defined similarly for in-sample factors $X$. Fig.\ref{fig-intro}(b, bottom and right) shows examples.  The bottom panel shows the limiting cases of $(\Delta m_{(1)}, \Delta m_{(2)},...)=(m,m,...,)$ that can enumerate all orders (ratio $\nicefrac{4!}{4!}=1$), the case of $(1,1,...,)$ that can enumerate only one ($\nicefrac{1!}{4!}=\nicefrac{1}{24}$) and an intermediary case. The figure also shows the corresponding squares (right), and thus effect observations, available across the resulting samples.

 
 We will demonstrate that effect generalizability assumes a particularly simple form with the following two sample gap conditions across time $t$, 

\begin{align}\label{eq-gap-conditions}
&P(x< u \, |\, \Delta U_{(0)}(t) = u_0, \Delta U_{(1)}(t) = u_1, ..., \Delta U_{(m)} (t)= u_m ) \\
&= P(x< u \,|\, \Delta U_{(0)}(t+m) = u_0, \Delta U_{(1)}(t+m) = u_1, ...) \tag{\textrm{fixed background}}\\
&= P(x< u \,|\, \Delta U_{(0)}(t+m) = u_{\delta 0}, \Delta U_{(1)}(t+m) = u_{\delta 1}, ...),   \tag{\textrm{random}} \nonumber
\end{align}

where $u_{0} \perp u_{\delta 0}$, $u_{1} \perp u_{\delta 1}$, etc. The first corresponds to the case of fixed (unobserved) effect backgrounds for a time period of at least $m$, and it can lead to generalizeable effects in fast $U{-}X$ systems. The second correspond to cases where effects are observed at time delays $\delta$, when assignment of unobserved confounders across gaps, $\pi_{\delta}$, no longer depends on the initial background, $\pi_0$. It is useful in in-sync and slow systems. 



  Confounding in samples following the conditions in Eq.(\ref{eq-gap-conditions}) can be expressed in the form

\begin{align}\label{eq-gap-linear}
U &= \frac{(q-1)}{q} \Delta U_{(1)} + \frac{(q-2)}{q-1} \Delta U_{(2)} + \ldots + \frac{(q-k)}{q-k+1} \Delta U_{(k)},
\end{align}

or, similarly, by saying that the variables $\Delta U_{(k)}$ form an (additive) Markov chain. The observation that order statistics under the previous conditions form a Markov chain was first made by Kolmogorov ~\citep{Renyi:1953ww}. If $U$ follows both exponential and uniform distributions, gap effects will have this additive property. While this is not always the case in non-experimental samples, decomposition of samples in subsets of effect observations with these properties will offer some advantages, in particular, in respect to sample error decomposition. More specifically than \textit{Proposition (\ref{prop-varordinal})}, Eq.(\ref{eq-gap-linear}) suggests that confoudness of effects in a square increases linearly across its columns, and is trivially associated with an ANOVA variance decomposition for effect variance in samples (\textit{Appendix \ref{sect-bayesian}}).

\begin{figure}
\centering
\includegraphics[width=1\linewidth]{figs/final/fig-2-ml-11.pdf}\\
\caption{ \textbf{(a)}  sampling sequences in a sample, $X=\{a,b,c\}$, with confounder partition $U=\{\{u_1\}, \{u_2,u3\}, \{u_4\}\}$,  \textbf{(b)}  normalized variation separation distance (distance to a Uniform distribution) of all backgrounds $\pi_0$ to other backgrounds after time $t$, vertical dashed lines mark limits on successive background randomness based on their number of inversions, Eq.(\ref{eq-inv}) (top panel shows the same without a test of stationarity on gap distributions $p$),  \textbf{(c)} randomization in graphical models with in-sample treatment $A$, other factors $W$, unobserved factors $U$, and outcome $Y$, \textbf{(d)} combinatorial relations in a square ($m=2$), circles are differences and gray circles are inter-unit factor value intersections (letters) between $x_0$ and other units.}\label{fig-intro2}
\end{figure}

\section{Sampling Treatment-Background Assignments}\label{sect-assign}


We will start with a definition for sampling sequences, assuming we can make repeated measurements for the observed variables in the DGP, $X$, but have no control over variables $U$. Eq.(\ref{eq-eff-def}) formulated a definition of effects for a single factor $a$, where backgrounds were $\Pi_n(Z/\{a\})$. It is generalized here for samples $X$, $\Pi_n(Z/X)$. These definitions rely on the following assumptions, which are somewhat typical,

\begin{assumption}
\begin{enumerate}
\item Constant DGP. The set of observed, $X$, and unobserved factors, $U$, in $Z=X\cup U$ do not change during the sampling period and there are no measurement errors. This leads to typical notions of irreducible variance and errors in samples ~\citep{Tibshirani:2001aa}.
\item Multiple Exposures. Effects remain constant under multiple exposures. While often assumed, this might be inadequate for interventions with continuous and cumulative effects, such as treatments involving dosages or complex interactions with exposures.  
\end{enumerate}
\end{assumption}

We also make the simplifying assumption,


\begin{assumption}\label{assumptions-2}
\begin{enumerate}
\item Non-strongly antagonistic unobserved interactions. When samples contain true causes \textit{Principle (\ref{prop-sep})} holds (\textit{Appendix.\ref{sect-omitted-corr}}). Many times, however, causes are unobserved but we still want to find in-sample surrogates that closely reproduce their root cause's effects. In this case, \textit{Principle (\ref{prop-sep})} might not hold when the effects of unobserved causes (i.e., observed in the same sample gap) strongly interact and can cancel each other. 


\end{enumerate}
\end{assumption}


Sampling sequences can be interpreted mathematically in several ways. We choose to think of them as permutations $\pi : b(t)$ of the set $K$ of integer sample gap indices. Imagine our effect sample observations to be discrete and equally spaced points in time. If an effect, sampled at time $t$, is next sampled at time $t'$, we write $b(t)=t'$.  If $t$ is an integer then let $t \bmod q$ denote the remainder when $t$ is divided by $q$.  Finally, we let $[t]_q$ be a shorthand for the set of possible remainders $\{0,1,2,...,q-1\}$ in this division.  We call each number $k$ in $[t]_q$ the sample \emph{gap assignment} for an unobserved factors $u \subseteq U$,  Eq.(\ref{eq-gaps}). Remember that permutations can have fixation points (positions whose order do not change). Each individual sample effect observation is consequently associated with a subset of unobserved factors $u \subseteq U$, Eq.(\ref{eq-gaps}), and we say,  in this case, that the effect is \emph{observed in the background of} $u$. Considering multiple sample factors, each permutation defines, in turn, distinct \emph{treatment-background} assignments.  Regularities across assignments define treatment-background \emph{mechanisms} in $U{-}X$ systems. 


\begin{defin}[Sampling Sequence]\label{def-background}
A sampling sequence over $X$ is a finite sequence $b$ of nonnegative integers $b(0), b(1),...,b(q-1)$ such that the mapping $b(t): K  \rightarrow K$ defined by $b(t) = t + b([t]_q)$ is a permutation of out-sample factors ('background' factors). A sampling sequence thus determines, for each time $t$, a surjective $U{-}X$ background-treatment assignment.
\end{defin}

 Fig.\ref{fig-intro2}(a) illustrates this concept in a sample $X_3=\{a,b,c\}$ with a fixed order statistic, $a < b <c$, and changing $U_4=\{u_1,u_2,u_3,u_4\}$ (slow sampling). The diagram shows the transition of backgrounds under repeated observations of effects with the fixed sample order $a<b<c$. Backgrounds are partitioned as $\{\{u_1\}, \{u_2,u_3\}, u_4\}$ in this example. Each arrow indicates the sequential mapping of unobserved factors (arrow labels) across sample gaps (intervals). See ~\citep{286326} for analogous combinatorial concepts.

 Effect backgrounds  $Z/X$ are not necessarily constant across repeated effect observations due to distinct frequencies, correlations and periodicities among $U$ and $X$ factors.  A key definition in this article is

\begin{defin}[EV-increasing sequence]\label{def-ev}
A finite sampling sequence $b$ of nonnegative integers $b(0) ... b(q-1)$, and its $U{-}X$ assignments, are \textbf{EV-increasing} if and only if the integers $[t+b(t)]_q$ are distinct.
\end{defin}

In Fig.\ref{fig-intro2}(a), for example, the top and bottom diagrams illustrate sequences that are not EV increasing, as effects for each variable effect are observed in only one background. The middle diagram is EV increasing with period $2$, as effects are observed in new backgrounds for that period, and all enumerable backgrounds for a fixed $Z/X$ are observed after time $2$.  A complementary property for sampling sequences is 

\begin{defin} [CF non-increasing sequence]\label{def-cf}
A finite sampling sequence $b$ of nonnegative integers $b(0) ... b(q-1)$, and its $U{-}X$ assignments, is confounding bias non-increasing, or \textbf{CF non-increasing}, if and only if the backgrounds $[t+b(t)]_q$ are the same across all sample factors $X$.
\end{defin}

In Fig.\ref{fig-intro2}(a), the top and middle diagrams are CF non-increasing. The bottom is CF non-increasing for a period of $2$. In \textit{Example (\ref{ex-extensions})}, the sequences $\mathbf{ab}u_1u_2u_3 \,, \allowbreak u_1\mathbf{ab}u_2u_3 \,,  u_1u_2\mathbf{ab}u_3 ,\, \allowbreak u_1u_2u_3\mathbf{ab}.$ are CF non-increasing, but so are any sequences with cyclic permutations of in-sample factors (i.e., subsequent rows of sample squares) under the condition that the unobserved background is fixed during the sample period. 

\begin{defin} [EV-CF sequences]
A finite sampling sequence $b$ of nonnegative integers $b(0), ..., b(q-1)$ is \textbf{EV-CF} if and only if it is both EV increasing and CF non-increasing. We say sample effect observations are 'generalizeable' under EV-increasing sequences, and  'unconfounded' in EV-CF sequences. 
\end{defin}

These sampling strategies can be described with modular arithmetic, $b(t+1) = c_1\times b(t) + c_2 \bmod q$, with, respectively, multiplicative ($c_1\gg 1$) or additive ($c_2\gg 0$) constants.  In abstract algebra, a group is a mathematical structure consisting of a set of elements and an operation (like addition or multiplication) that satisfies certain properties. In modular arithmetic, the integers coprime (relatively prime) to $q$ from the set $\{0,1,\dots,q-1\}$ of $q$ non-negative integers form a group under multiplication, called the multiplicative group of integers modulo $q$. Equivalently, the elements of this group can be thought of as the congruence classes, 'residues' modulo $q$, that are coprime to $q$.

Sequences generated by recursive multiplication ($c_1\gg 1$) exhibits more non-linear and less predictable behavior compared to addition ($c_1\approx 1$).  One group-theoretic property of interest is closure, which states that the result of performing the operation on any two elements of the group is still an element of the group. In the context of addition modulo \( q \), addition satisfies the closure property. This means that for any two residues \( i \) and \( j \) modulo \( q \), \( i + j \) is still a residue modulo \( q \).  However, when we consider multiplication modulo \( q \), closure is not guaranteed. Not every pair of residues modulo \( q \) has a product that remains within the set of residues modulo \( q \). This lack of closure allows multiplicative recursions ($c_1\gg 1$) to include numbers in different residue classes modulo \( q \), and, at the same time, introduce more complexity and non-linearity compared to addition. This observation is often employed by random number generators ~\citep{10.5555/270146} as means to generate random numbers. 


\subsection{Background Enumeration} 


We first consider fast sampling, the case where $\Delta_{(1)} m > \Delta_{(1)} q$ in specific sampling periods. In this case, in-sample effects are observed for all factors in constant backgrounds. The constant $c_2$ describe 'phase' differences that, when $c_2$ and $q$ are co-prime, generate samples with the cyclic permutations of $X$ (individual square rows).  Systems under fast sampling are EV-increasing for a constant sampling rate when $c_1/q$ is integral, 

\begin{defin}[$U{-}X$ Effect Background Enumeration]\label{def-enum}
For a fixed background $Z/X$ (and a constant sampling rate), sequences $b(t+1) = c_1\times b(t) + c_2 \bmod q$ for  $\nicefrac{c_1}{q}$ integral and $\text{gcd}(c_2, q) = 1$ are EV increasing and CF non-increasing, where $\text{gcd}(c_2, q) = 1$ indicate that $c_2$ and $q$ are co-primes.
\end{defin}




 Consider, for example, $c_1/q=2$. For a fixed gap assignment $k$, it takes $2^t \times k (\bmod 2n-1) = k$ time for $u_k$ to get back to its original position ~\citep{Diaconis:1983uo}, and for all background factors to be enumerated\footnote{as result of the recursion $(2(2...(2k(\bmod 2n-1))...)))$ for each gap position $k$.}. This  can be restated as

\begin{align}\label{eq-convolution}
2^t \ (\bmod \ 2q-1) = 2^t \ (\bmod \ q-1) =1,
\end{align}

for a initial position $k=1$. This condition is known to only 'circulate' the set of treatment-background assignments, $b(t)$, and lead to no randomization ~\citep{Diaconis:1983uo,diaconis2023mathematics}. For poset chains $\{G(X),G(U)\} =\{\;  \mathbf{a} < \mathbf{b}, \;\;u_1 < u_2 < u_3 <u_4\;\}$,  for example, where $q=4$ and $c_1=2$, linear extensions of type $\mathbf{a}u_1u_2\mathbf{b}u_3u_4$ can generate EV-increasing sequences with a constant sampling rate, unlike $\mathbf{a}u_1u_2u_3\mathbf{b}u_4$.

\subsection{Background Randomization} 



We next consider the case of in-sync and slow sampling, $\Delta_{(1)} m \leq \Delta_{(1)} q$. Both in-sync and slow sampling can be seen as allowing the multiplicative and integral constant $c_1$ to take random values. In the in-sync case, $c_1$ takes approximately common $\Delta_{(k)}q$ values across all $k$, $0<k\leq m$. It then describes background-treatment assignments where the $q$ confounders are equally distributed across sample gaps, $\Delta_{(k)}q\times m \approx q$, across time. In \textit{Example (\ref{ex-extensions})}, we exemplified how sample effect estimation depends on the extensions between posets $G=\{G(X),G(U)\}$. We then described them as partitions of $U$ induced by sample gaps, Eq.(\ref{eq-gaps}). We now describe this relationship with a multinomial sample gap distribution, $p = \{p_1,p_2,...,p_m\}$, counting the number of distinct confounders assigned to a gap $k$, $\Delta_{(k)} q$, after multiple sampling cycles.
 



Imagine we keep $p$ constant from time $t=0$ to $\delta t$ while allowing $\Pi_n(Z / X)$ to change.  Can we say anything about whether the backgrounds $\pi_0$ and $\pi_{\delta t}$ are random-like (i.e., the realization of $U$ in the first is as 'separated' to the second as independent samples of a uniform distribution), $\pi_0 \perp \pi_t$? We can describe this relationship more precisely by fixing an initial permutation $\pi_0$ and counting the number of inversions, or pairs of elements that are in the reverse order, from $\pi_0$ at the subsequent time $t$. Fix $p$ and two backgrounds $\pi_0, \pi_1  \in \Pi_n(Z / X)$ separated by only one inversion, for example. The likelihood of reaching $\pi_1$ from $\pi_0$ after one time step is high in this case. With increasing inversions in $\pi_0$, there should be some period $\delta t$ where that likelihood becomes comparable to that of a uniform distribution, where the expected number of inversions generated at each time is $\binom{m}{2}$.  The expected number of permutation inversions, $I(\pi_0,t)$, required for that condition ('randomization') with a fixed distribution $p$, initial background $\pi_0$, and under i.i.d. sampling ~\citep{Fulman:1998uz} is 

\begin{align}\label{eq-inv}
\Expect [\,I(t \,|\, p, \pi_0)\,] &= \frac{1}{2} \times \binom{m}{2} \times \Big(1-\Big[p_1^2 + p_2^2 + ... + p_m^2\Big]^t \Big),
\end{align}


 after time $t$. With a $p$ close to $\{1/q,1/q,1/q,...\}$, randomization happens quickly, as the term within the right-most parenthesis in Eq.(\ref{eq-inv}) converges quickly to $1$, with an increasing $t$.  The choice of a background $\pi_1$ from $\pi_0$ is then as likely as any other possible background. For the case of $p=\{0.5,0.5\}$ the equation reduces to

\begin{align}\label{eq-inv-binom}
 \Expect [\,I(t \,|\,\pi_0)\,] &= \binom{m}{2} \times \Big(1-\frac{1}{2^t}\Big),
\end{align}

which underlies the known cut-off limit for markov chains ~\citep{Diaconis:1996vh,bayer1992trailing}. As consequence, for $m=2$ and random treatment-background assignments ($p_1{=}p_2{=}0.5$ and $U{-}X$ sampling i.i.d.), effect generalizability requires in the order of $\mathcal{O}(log_2t)$ effect samples. This is a stochastic generalization of \textit{Definition (\ref{def-enum})} and the previous case of $c_1=2$ on  fast $U-X$ systems, $\nicefrac{\Delta_{(k)}}{q} \sim Bin(q, 0.5)^{-1}$. Another trivial example has probability mass $p=\{1,0,0,...\}$ concentrated in a single gap, which, as expected, generates no inversions. This leads to the limit $|\Pi_n| = 0$ which describes samples with no EV increasing sequences. A sample with $p$ close to $p=\{1/q,1/q,1/q,...\}$ ($q\geq m$) is capable of asymptotically enumerating $|\Pi_n| = q!$ orders of unobserved causes in the least time $t$, Eq.(\ref{eq-gapprob}). This condition is associated with $U{-}X$ independence, and, departures from this condition with more restrictive requirements on sampling times or sample sizes. The number of EV-increasing permutations, for a given $t$, is decreasing as we depart the last limit (random $U$) and approach the previous (deterministic $p=\{1,0,0,0,...\}$).   For intermediary and more typical scenarios,  what matters is the particular distribution $p$ of confounders across a sample's gaps, Eq.(\ref{eq-inv}). Notice that $\binom{m}{2}$ in Eq.(\ref{eq-inv}) is the number of effects observation in a single square. 


The relationship between inversions and randomness  is a deep, and somewhat unresolved, issue in mathematics ~\citep{bayer1992trailing,diaconis2023mathematics}. With an assumption of stationarity of $p$ (\textit{Sect.\ref{sect-backrandom} \textit{\nameref{sect-backrandom}}}), Eq.(\ref{eq-inv}) suggests under what conditions (and time after $\pi_0$) we can say that \textit{any} treatment-background assignment selected would be chosen with equal probability.
  
  \begin{defin}[$U{-}X$ Effect Background Randomization]\label{def-random}
Under non-fixed backgrounds $Z/X$ and stationary $p$ (i.i.d.), sampling sequences $b(t+1) = c_1\times b(t) + c_2 \bmod q$ and  effect treatment-background enumeration can become asymptotically random, and thus EV-increasing with high-likelihood. For an initial effect background $\pi_0$, sample sizes required for this condition depend on the treatment-assignment distribution $p$, and the number of inversions from $\pi_0$ the distribution $p$ can generate, Eq.(\ref{eq-inv}).
\end{defin}

Fig.\ref{fig-intro2}(b) illustrates this limit for the simulations in \textit{Sect.\ref{sect-exp} \nameref{sect-exp}}, which include samples with unobserved, correlated and non-stationary factors. The variation separation distance (y-axis) is a measure of randomness formulated as a distance to an uniform distribution ~\citep{bayer1992trailing,Diaconis:1996vh}. The panel shows the distance between a fixed background $\pi_0$ and the subsequent in samples where $p$ is stationary (according to a Kendall-tau test, permutations and inversions are exhaustively enumerated and counted using the tests and procedures described next, distances were normalized to their per-factor range for display). The upper panel shows the same for other samples where $p$ is not stationary (according to Kendall-tau). These combinatorial considerations suggest limits where some guarantees on effect background randomization  can be made. In systems with $q\gg m$, they indicate that effects will be observed in new backgrounds after each formulated period and, thus, that effects are observed under EV-increasing sampling sequences.

 The proposed condition is only valid for a stationary $p$ and initial $\pi_0$.  We associate EV-CF sequences, instead, with $m$ sampling sequences with the same properties but observed under $m$ in-sample cyclic permutations (i.e., with period $m\times \delta t$). The consequence of the later is that effects for all factors are observed across the same backgrounds for all in-sample factors $X$. These i.i.d. and permutation inversion conditions are further discussed in \textit{Sect.\ref{sect-backrandom} \nameref{sect-backrandom}}. In the context of effect estimation, Eq.(\ref{eq-inv}) relates slow sampling to ideal in-sync sampling (as visualized by squares) and the conditions under which they can lead to samples with generalizable effects. They implement the stipulated gap distributions in Eq.(\ref{eq-gap-conditions}) and lead to the proposed solutions here for causal effect estimation with increasing generalizability,
 

\begin{prop} 
Counterfactual background effect enumeration, Definition (\ref{def-enum}), is EV increasing in fast $U{-}X$ systems, and background randomization, Definition (\ref{def-random}), is EV increasing in in-sync and slow systems.
\end{prop}



 
 


\section{Square Enumeration and Gap Randomization Tests}\label{sect-backrandom}



Both the randomization conditions in the last section and \textit{Principle (\ref{prop-sep})} are related to permutation inversions.  Inversions appear in standard tests for i.i.d. sampling ('stationarity') in ordinal data, such as in  Kendall-tau and Man-Whitney tests. Any pair of  observations $(x_i,y_i)$ and $(x_j,y_j)$, where $i < j$, are said to be 'concordant' if the sort order of $(x_i,x_j)$ and $(y_i,y_j)$ agrees: that is, if either both $x_i>x_j$ and $y_i>y_j $ holds or  both $x_i<x_j$ and $y_i<y_j$; and be 'discordant' otherwise. The number of discordant pairs between $x$ and $y$ is equal to the number of inversions that permutes the $y$-sequence into the same order of $x$. The statistic can also be seen as a generalised measure of a classifier's separation power for more than two classes ~\citep{Hand:2001us}. 

 We now formalize the previous conditions for effect background randomization in non-experimental samples. They are fulfilled by any subsample $C$, $C \subset X$, passing the following statistical hypothesis tests over inversions, 


\begin{align}\label{eq-sq-conditions}
  \begin{cases}
 \; \mathbb{I}_{C}\Big[ \, \Delta^2 y_{(k)} (X{=}x)\,\Big]_t=0, \\
  \; \mathbb{I}_{C}\Big[\, \Delta y_{(k)} (X{=}x)\, \Big]_t  > 2\times\Big(p_1^2 + p_2^2 + ... + p_m^2\Big)^{-t} 
 \end{cases},
\end{align}


where $\mathbb{I}[.]_{C}$ are permutation inversion counts over ranked gap outcome differences-of-differences or differences in the subsample $C$, $|C|\geq 2$ and $n>0$. The first test is a (Kendall-tau) test of stationarity for $p$, and the second test uses Eq.(\ref{eq-inv}) and \textit{Assumptions (\ref{assumptions-2})}. Together, they formalize the previous condition that factors $U$ should be changing for a period above the limit formulated by $Eq.(\ref{eq-inv})$, while $p$ should remain stationary in the same period. They are the out-of-sample analogue of the effect differences illustrated in \textit{Example (\ref{ex-diffs})}. \textit{Principle (\ref{prop-sep})} is also related to inversions and says that confounders, causes and spurious factors have distinct effect variances under increasing inversions (i.e., the number of times a confounder is expected to be sampled before its root cause). According to the principle, the observation of one inversion allows for the separation of a single cause-confounder pair (\textit{Appendix.\ref{sect-cf},\ref{sect-omitted-corr}}). 

We can also generalize the previous conditions to samples that are not sequentially sampled. This is done by enumerating squares for a specific sample unit $x_0$ using inter-unit combinatorial properties. In fully observed samples, $q=0$, square enumeration ensures the applicability of \textit{Principle (\ref{prop-sep})}. In samples with unobserved factors, $q>0$, enumeration with the additional conditions over gap inversions in Eq.(\ref{eq-sq-conditions}) offer analogous advantages. Much like we can observe single factor differences in samples, and their effects,  we can observe factor permutations and their effects, by combining across-unit differences and their effects carefully. The panel in Fig.\ref{fig-intro2}(d) illustrates this process for two factors ($m=2$). It illustrates the differences (circles) and intersections (gray), from $x_0$, that characterize each cell in a square of size $2$.  We say that by observing differences $a$, $b$ and $ab$ from $x_0$, we 'observed' a permutation for these factors. That's because these differences - $(x_0{-}x_1) = \{a\}$, $(x_0{-}x_2) = \{b\}$ and $(x_0{-}x_3) = \{ab\}$ - allow us to recover the effects of permuting factors $\{a,b\}$ (conditional on $x_0$), which are not directly observed in the sample. Notice that not all sample pairs with differences $\{a\}$ and $\{ab\}$ would belong to the second square column, as there are also conditions on their intersection, $(x_0{-}x_1) \cap (x_0{-}x_3) = \{b\}$. It is the fulfillment of both  combinatorial conditions that allow us to take them as (conditional) effect observations. The square full combinatorial structure generalizes this to $m > 2$ and uses the $2^m$ $m$-way differences potentially available in samples.  Further combinatorial, computational and statistical aspects of this enumeration process are described in \textit{Appendix.\ref{sect-cf},\ref{sect-omitted-corr},\ref{sect-enumeration}}. 

Fig.\ref{fig-intro}(a) summarizes the proposed solution for effect estimation, where square enumeration ensures the inversions required by \textit{Principle (\ref{prop-sep})} for in-sample factors, and tests over their gaps, Eq.(\ref{eq-sq-conditions}), for out-of-sample factors. For sets of units (and thus mutual effect observations) where these combinatorial and gap conditions hold simultaneously, variance in effects can become asymptotically indicative of the status of in-sample factors as causes, \textit{Principle (\ref{prop-sep})}, their effects, Eq.(\ref{eq-eff-def}), and their effects generalizability, Eq.(\ref{eq-ev-def}). Variance becomes, in resulting samples, a measurement of \textbf{'noise' for individual effect observations} that also takes causal principles into consideration ~\citep{F.-Ribeiro:2022tm}.

 



\section{Related Work}\label{sect-related}

Non-parametric tests (like permutation tests) can often only be used with strict assumptions on sample completeness. The previous discussions suggested the need to permute (unobserved) effect backgrounds when estimating effects. According to our view, due to the frequent partial observability of samples, the challenge for estimation then becomes how to establish in-sample conditions indicating when backgrounds (1) have not changed, or, (2) that they have changed enough such that any association between treatment and backgrounds of two effect observations or units have nearly vanished.  We suggested that a way to establish such conditions, when the number of unobserved factor is large,  is to seek assurances of randomness from established theory, such as from Combinatorial Randomization theory ~\citep{Diaconis:1996vh,diaconis2023mathematics}. 


Our main goal has been to formulate a \emph{quantifiable} model for the generalizability of effect observations across populations. A lack of understanding about the issue  creates artificial issues that surface once and again in non-parametric experimental studies, observational causal estimation, and black-box variable importance attribution. For example, Ribeiro et al. ~\citep{F.-Ribeiro:2022tm} showed that lack of sample variation explains one of the most heated methodological debates in Economics. Grimmer et al.  ~\citep{Grimmer:2020aa} mentions how the presence of 'Stan Lee' in the credits of all Marvel franchise movies, as factor when estimating effects on movie success, leads celebrated algorithms like the Deconfounder ~\citep{Wang:2019aa} to drastically overestimate the factor's effect.  More than any particular algorithm, what seems missing is a sound way to quantify the EV of effect estimates, and a better understanding of the issue altogether. 

Concerns about robustness and predictive performance can, in part, be seen as the motivation for more recent views of causality based on {model-performance} invariance ~\citep{Peters:2016aa,Magliacane:2017aa,Buehlmann:2020aa}. This is typically the invariance of a black-box model's performance - and not of an effect, or, \textbf{difference in outcomes}, in a population, as taken here. The unit of analysis here, $\Delta y$, is much simpler (being an observation and not an estimate).  Furthermore, on the present view, it is not the {minimum variance} of effects that matters, but the minimum variance of effects in face of the {maximum variance} of extraneous factors.  The approach here formulates and addresses the latter conditions.  Finally, we believe the widespread success of the counterfactual definition is largely due to its simplicity, and  simplicity has been a central guideline in the present work. 


The issue of the Extenal Validity (EV) of effects has been addressed parametrically in ~\citep{Pearl:2014ug,6137426,NEURIPS2020_7b497aa1}. Authors assume a known 'selection' variable set $S$ that 'may represent all factors by which populations may differ or that may threaten the transport of conclusions between populations' when training parametric models on non-experimental data. They then, assuming causal graphs for all other variables are known or inferred for both populations, propose graphical conditions for EV based on Pearl's graphical calculus.  Assuming all this prior knowledge is known, the proposed conditions may give  assurances on the transportability of 'observational findings' (parameter estimation). Often the previous differences are not known. We take a more practical starting point, where every variable is potentially a relevant difference, no independence relations are assumed, and all results follow from in-sample effect differences and probabilities. We assume that subsets of 'population' factors are observed ($X$) but another set of factors ($U$) is completely unobserved, as typical in supervised training tasks. We provide a {quantifiable} measure of EV (not a set of ideal graphical and independence conditions) for samples. In our view, the EV of effects is an uncertainty measure, Eq.(\ref{eq-ev-def}), and not an isolated issue. The solution is formulated at the level of individual effect observations, non-parametrically, and in the counterfactual framework. At the same time, it is difficult to think of issues of external validity separate of issues of selection bias (which is the central issue in scientific research and Experimental Design) ~\citep{Cox:2009tn,Montgomery:2001aa} and randomization.  We consider whether connections to Combinatorial Randomization might be helpful. The alternative combinatorial model proposed allow us to discuss critical issues - difficult to address, even if only theoretically, in the previous - like in-sample and out accuracy trade-offs, selection bias, prediction-interpretability tradeoffs, empirical results, and sample size requirements.

  \subsection{Randomization and Parametric Inference}


We will use the notation often employed in Structural Causal Models (SCM) to relate the previous concepts to research in parametric effect inference. In SCM models, variables are typically specified by an unknown function of a set of endogeneous variables and an exogeneous error. We have, similarly, specified DGPs $Z=\{X,U\}$ with a collection of exogenous factors $U$, in addition to the observed $X$. Factors in $U$ are unobserved and not affected by $X$, but can affect $X$. More specifically, a SCM specifies a collection of functions $f_a$ indexed by endogenous variables, $f =(f_a : a,b,c,...)$. The SCM also calls for the specification of the relation between the random variables $U$ and $X$. This is done through a set of functions $f=(f_W,f_a,f_y)$, $A$ is a random indicator for a treatment $a$, $W$ are the other (non-treated) in-sample factors, $W= X-\{a\}$, and $y$ is the outcome of interest. The values of $W$, $A$, and $Y$ are then deterministically assigned by $U$, according to ~\citep{van2011targeted}

\begin{align}\label{eq-scm}
W=f_W(U_W)\\
A=f_a(W,U_a\nonumber )\\
Y=f_y(W,A,U_y) \nonumber
\end{align}

which defines a treatment-background assignment mechanism (without assumptions about $f_W$,$f_a$, and $f_y$).

Randomized controlled trials (RCTs) are one type of controlled experiment where subjects are randomized to receive a specific treatment. For example, when each subject is assigned to a treatment based on the flip of a coin. In this case, differences between the outcomes of treatment and control groups indicate the treatment's effect, as all other factors would be \emph{balanced} (free of selection bias), up to a random error. In an RCT, the treatment assignment mechanism, $P(A = 1 | W)$, is known (e.g., constant and $0.5$ in the coin flip). This is different from non-experimental data  where treatment, or treatment exposure, is not assigned. A RCT can be, nonetheless, written in SCM notation as in Eq.(\ref{eq-rctw}-\ref{eq-rcta},\ref{eq-rcty}),

\begin{align}
W &= f_W(U_W), \label{eq-rctw}\\
A &= f_a(U_a), \; A \perp W \label{eq-rcta}\\
&\textrm{(a)\;} \begin{cases}\label{eq-rctthis}
 &= f_A(U_a \, |\, \Pi_n(Z/\{a\})\\
 &= f_A(U_a \,|\, \pi_0, \delta),\; \pi_0  \perp \pi_{\delta}
\end{cases}\\
Y &= f_y(W,A,U_y). \label{eq-rcty}
\end{align}

The only difference between Eq.(\ref{eq-scm}) and Eq.(\ref{eq-rctw}-\ref{eq-rcta},\ref{eq-rcty}) is the structural equation for $f_a$ in Eq.(\ref{eq-rcta}). The baseline variables $W$ play no role in the generation of $A$. These assumptions can also be illustrated with causal graphs, Fig.\ref{fig-intro2}(c). Since $a$ is randomized, this implies that $U_a$ is independent of $U_y$ and $U_W$, and, in this case, no arrows connect $U_a$ to $U_y$ and $U_a$ to $U_w$ in the graphical model. Eq.(\ref{eq-rctthis}) reviews the conditions on sample gap distributions proposed in this article and in Eq.(\ref{eq-gap-conditions}).

With random assignment of backgrounds to fixed treatments, as the sample size $n$ approaches infinity, by the law of large numbers, the average characteristics of treatment and control groups, $P( Z-\{a\} < a )$ and $P( Z-\{a\} >a )$, converge to population averages. This can be written as

\begin{align}\label{eq-randomization}
\lim_{{n \to \infty}} P( Z-\{a\} < a ) = P( Z-\{a\} >a ).
\end{align}

With convergence, the potential bias introduced by correlations in observed-unobserved ($U-X$)  is mitigated. On average, any unobserved or uncontrolled factors that could potentially bias effect estimates will be equally distributed between the two effect observation orders, as required by \textit{Principle (\ref{prop-sep})}. In a square, this is achieved by the stipulated requirements for the randomization of $Z-\{a\}$ factors, Eq.(\ref{eq-sq-conditions}). 








\subsection{Non-parametric Effect Estimation}\label{eq-nonparam}

The way both treatment assigments $A$, Eq.({\ref{eq-rcta}}), and effects $Y$, Eq.({\ref{eq-rcty}}), are derived in the present approach are related to other non-parametric causal effect estimators. Targeted learning is a semi-parametric approach  that starts by specifying the mechanism by which individuals were assigned to different treatments. It does not assume random assignments.  Specifically, taking the effect of a factor $a$ as the target parameter, it's two main components are a treatment assignment mechanism \( Q \), and an empirical distribution \( P_w \). The first defines the probability distribution of receiving a particular treatment (e.g., the conditional probability of treatment given observed factors) and the second is the observed distribution of $W = X-\{a\}$ considering only those individuals who received the specific treatment (i.e., the distribution of the observed data for individuals who were treated according to a particular mechanism).  The targeted learning estimator \( \hat{\Delta y(a)}_n \) is then obtained by solving the following optimization problem:

\[ \hat{\Delta y(a)}_n = \arg\min_{\theta} \left\{ \frac{1}{m-1} \sum_{w \in X-\{a\}} L_{\theta} \left(\Delta y(a), Q, P_w(a)\right) \right\} \]

Here, \( L_{\theta}  \) is the loss function over hyperparameters $\theta$ that quantifies the discrepancy between the true parameter \( \Delta y(a) \), the data-generating distribution \( Q \), and the empirical distribution \( P_w \) induced by a specific treatment assignment. The optimization is typically performed using ensemble or black-box machine learning techniques. The optimization thus aims to find hyperparameters \(\theta\) that minimizes the loss across all treatment assignment mechanisms, resulting in an estimator that is robust to different ways treatments may have been assigned in the non-experimental sample. The approach here has the same goal, but relies on a theoretical connection between treatment-background assignments, in one hand, and their relation to generalizability and confoundness of effects on the other. The theory provides an answer to when specific effect observations can be said to have the later properties, conditional on a fixed mechanism (as identified by sample gap statistics and tests). We seek to demonstrate that, under specific mechanisms, samples contain effects that are externally valid.  The approach holds a similar relationship to parametric methods that employ black-box techniques across time to generate treatment-control pairs for effect estimation \citep{10.1162/rest_a_00760,10.1257/jel.20191450}.

In respect to outcome functions, Eq.({\ref{eq-rcty})}, a general non-parametric strategy is to condition outcomes on all in-sample factors and take their difference in means in each resulting stratum.  This leads to an estimate of the effect of treatment as the average over all strata which is significantly more efficient than an unadjusted treatment effect estimate. That's because the Law of Total Probabilities allows such methods to control (in-sample) confounding without the assumption that effects are constant across strata.  If $q=0$, the g-formula ~\citep{Keil:2014uc}  implements this idea.  For example, an outcome could be expressed as

\begin{align}\label{eq-gformula}
P(y=1) = \sum_{W=X-\{a\}} P( y=1 |  W = w) \times P( w)
\end{align}

which indicates that we are summing over each possible value of $W = X -\{a\}$, and $P(W = w)$ is the probability that $W$ takes on the value $w$ in the reference population when only $a$ is left to vary. This adjusted estimator of the effect of $a$ is a nonparametric maximum likelihood estimator for $Y$ for a single treatment ~\citep{Keil:2014uc,van2011targeted}. Drawbacks are the requirement that $q=0$ and the loss in performance in strata with a small number of units. Analogously, in Eq.(\ref{eq-eff-def}) effects are conditioned on all their backgrounds (where order is also taken to matter). The strategy in Eq.(\ref{eq-gformula}) is often combined with conditional independence tests to identify plausible in-sample confounders. While we presently favor simplicity to highlight theoretical considerations and unobserved confounding, elements of both these approaches (such as further in-sample conditional independence tests of the g-formula and the supervised plug-in flexibility of targeted learning) could be further combined with the proposed.

\section{Sample Sizes}\label{sect-power}

Non-experimental samples may contain subsamples that are simultaneously unbiased and predictive. Finding such subsamples shifts the requirement from one of carrying out an experimental intervention to one of accruing sufficient sample sizes across required combinatorial conditions.  The expected sample size necessary to sample a square, when sample factors have different frequencies, is related in a simple way to its ordinal statistics. The asymptotics for recommended sample sizes, $\tilde{n}$, are, in particular,

\begin{equation}\label{eq-n}
2^{m-1}P[\, X_{(m-1)}\,]^{-1}, \quad \text{and,} \quad  \Big(2^{m-1}P[\,X_{(m-1)}\,]^{-1}\Big)\times \phi,
\end{equation} 
 
  for the first and many squares scenarios, and where $\phi= 1.6180...$ is the golden ratio (\textit{Appendix.\ref{sect-power-app}} has a full proof). The number of squares in observational data are, this way, determined by its rare factors, which demand samples with increasing sizes to match the number of permutations in their balanced counterparts (i.e., where factors have equal frequencies). 
  


\begin{figure}
\centering
\includegraphics[width=0.9\linewidth]{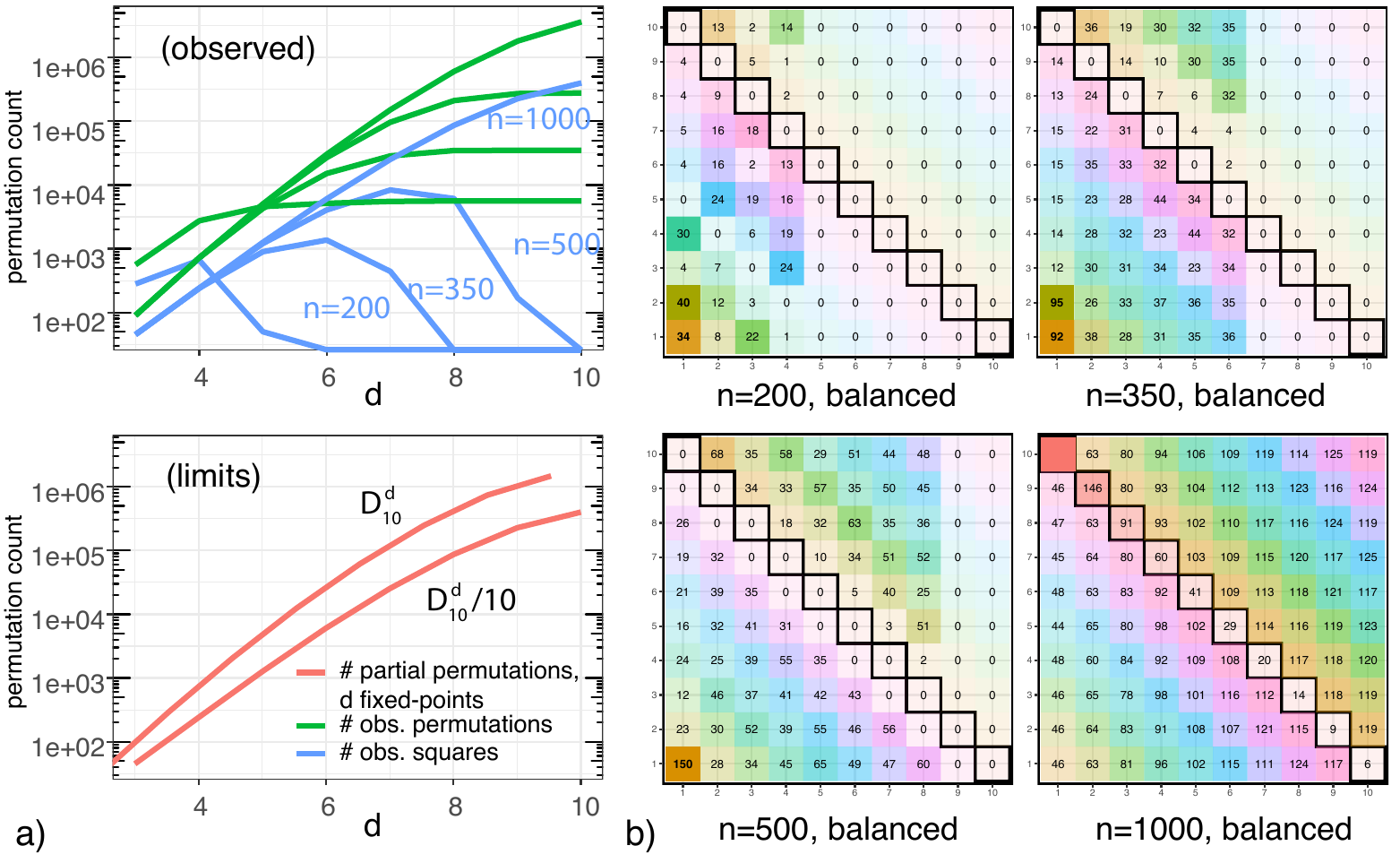}\\
\caption{ \textbf{(a)} observed permutations and limits (m{=}10, balanced, log-scale); \textbf{(b)} square sample-unit histograms, increasing $n$.}\label{fig-count}
\end{figure}

\section{Experiments}\label{sect-exp}

We consider performance of supervised, black-box explaining, and effect inference algorithms across a range of simulations and a real-world application. We thus study three distinct prediction {problems} in held-out samples with $n$ observations. The \textbf{tasks} are to predict

\begin{itemize}
  \item[$-$] $y_i \in \{\minus 1,\plus1\}$, from $x_i \in \{\minus 1,\plus 1\}^m$, $\forall  i \in \{1,...,n\}$, \quad (\emph{supervised})
  \item[$-$] $\Delta y(x_{ij})\in \mathbb{R}$, from $x_{ij}{=}x_i{-}x_j \in \{\minus 1,\plus 1\}^m$, $\forall (i,j)\in\{1,..., n \}^2$,  \quad (\emph{counterfactual})
  \item[$-$] $\Delta y(a) \in \mathbb{R}$, from $x_i \in \{\minus 1,\plus 1\}^m$,  $\forall  a \in X$.  \quad (\emph{effect or importance})
\end{itemize}

These correspond to out-of-sample prediction of outcomes, prediction of intervention effects, and marginal factor effect estimation (evaluated against the ground-truth in simulations).  The first is the central problem in supervised Machine Learning, the second is related to effect estimation with counterfactual and matching estimators \citep{Morgan:2007aa}, and the third to both causal effect estimation \citep{RefWorks:doc:5911ec76e4b0ac17f7d9f458} and supervised explainability ~\citep{10.1613/jair.1.12228}. Effect estimation is, in the present perspective, an extreme omitted-variable problem. The problem becomes easier with decreases in the number $q$ of unobserved factors and the proposed framework can thus also be used to study the performance of generic samples and models.

 Simulations follow common binary generative models ~\citep{Chatton:2020aa}, with $m$ Binomial observed factors, $x \in \{\minus 1,\plus1\}^m$, $q$ unobserved, $z \in \{\minus 1,\plus1\}^q$, and sigmoidal outcomes, $y = \text{sigmoid}(\sum_{a\in X} x(a)\Delta y(a)) \in \{\minus 1,\plus1\}$ (as in logistic and many categorization models). With $X_m$ the set of $m$ observed factors and $U_q$ the set of $q$ unobserved, the \textbf{simulated cases} are

\begin{itemize}
\item[$-$] stationary and equiprobable factors with constant effects,\quad (\emph{balanced})\\ $p(a) =0.5, \: \Delta y(a) = 1, \: \forall  a \in X_{10}$ and $\forall u \in U_{3}$, 
\item[$-$] non-stationary and factors with distinct probabilities and effects, \quad(\emph{unbalanced})\\
$p(a) \sim \mathcal{U}([0,1]), \: \Delta y(a) \sim \mathcal{U}([0,1]), \: q \sim \mathcal{U}([3,20]), \: \forall  a \in X_{10}$ and $z \in U_{q}$,
\item[$-$] non-stationary and correlated factors,  \quad (\emph{correlated}, $\times 3$)\\
$p(a) \sim \mathcal{U}([0,1]), \: \Delta y(a) \sim \mathcal{U}([0,1]), \: q \sim \mathcal{U}([3,20]), \: \rho (a,b)=\mathcal{U}([0,\{0.1,0.25,0.5\}]), \:b = next(a),  \: \forall  a \in X_{10},$ and $\forall z \in U_{q}$,\\
 where $next(a)$ is $a$'s subsequent letter in lexicographic order,  
\end{itemize}
 
 For the last two cases (\emph{unbalanced} and \emph{correlated}), we also divide the sampling period into 10 time sub-periods, where at the beginning of each period the number of unobserved factors and parameters are re-sampled, while in-sample effects remain constant.
 
For $m$ factors and $d$ fixed-points, the number of permutations, derangements (permutations without overlaps), and partial permutations (permutations with $d$ positions fixed) ~\citep{Hanson:1983aa} are, respectively, 

\begin{equation}\label{eq-limits}
m!, \quad D_m = m! \sum_{d=0}^m \frac{(-1)^d}{d!}, \quad \textrm{and}, \quad D_{m}^d =  \binom{m}{d}D_{m-d}.
\end{equation}

 
The last number indicates that, to form a partial permutation, we select $d$ unique factors to fix and be organized in $\binom{m}{d}$ ways, each with $D_{m-d}$ possible disjoint orderings of the non-selected factors. Fig.\ref{fig-count}(a) shows the number of {observed} permutations present on increasingly larger samples (\emph{balanced}).  Each (green) curve corresponds to the number of permutations in samples of size $n=\{200,350,500,1000\}$ (blue curves show the number of squares). The lower panel shows combinatorial limits (red) for the number of partial permutations and squares according to Eq.(\ref{eq-limits}), $m=10$.  The figure illustrates, in particular, the relationship between sample sizes $n$ and the subset of permutations $\Pi_n(Z)$ that effects for factors $Z$ can be asymptotically observed under\footnote{since $q\leq 20$ in these illustrations all permutations in a sample can be enumerated with the approach in \textit{Appendix.\ref{sect-enumeration}}}, for each number $d$ of fixation-points, $d\leq m$. The relationship between the DGP and enumerable permutations or their 'speed of enumeration' was, for example, formulated in Eq.(\ref{eq-gapprob}) and illustrated in Fig.\ref{fig-intro}(b). We expect these quantities to impact sample biases and predictive performance, Eq.(\ref{eq-eff-def}), irrespective of the DGP. The same is shown as histograms of square member counts,  Fig.\ref{fig-count}(b). At $n{=}200$, no differences of size larger than $4$ are observed. A full square of size $10$ is completely observed only at $n{=}1024$ (bottom-right) in the \emph{balanced} case, and $n\approx 20K$ in the \emph{unbalanced} case. This was anticipated by Eq.(\ref{eq-n}) and follow from simple sample ordinal statistics. The first sample size is $\tilde{n} = 2^{9+1}=1024$. The latter sample has rare factors, $\Expect [min_{a\in X}(p(a))]= 0.025$, which leads to an approximate sample requirement of $\tilde{n}=2^{9}/0.025= 20480$. Similar sample size recommendations will be indicated for each of the increasingly complex simulated and real-world cases considered below.

\subsection{Outcome Prediction}

\begin{figure}
\centering
\includegraphics[width=1\linewidth]{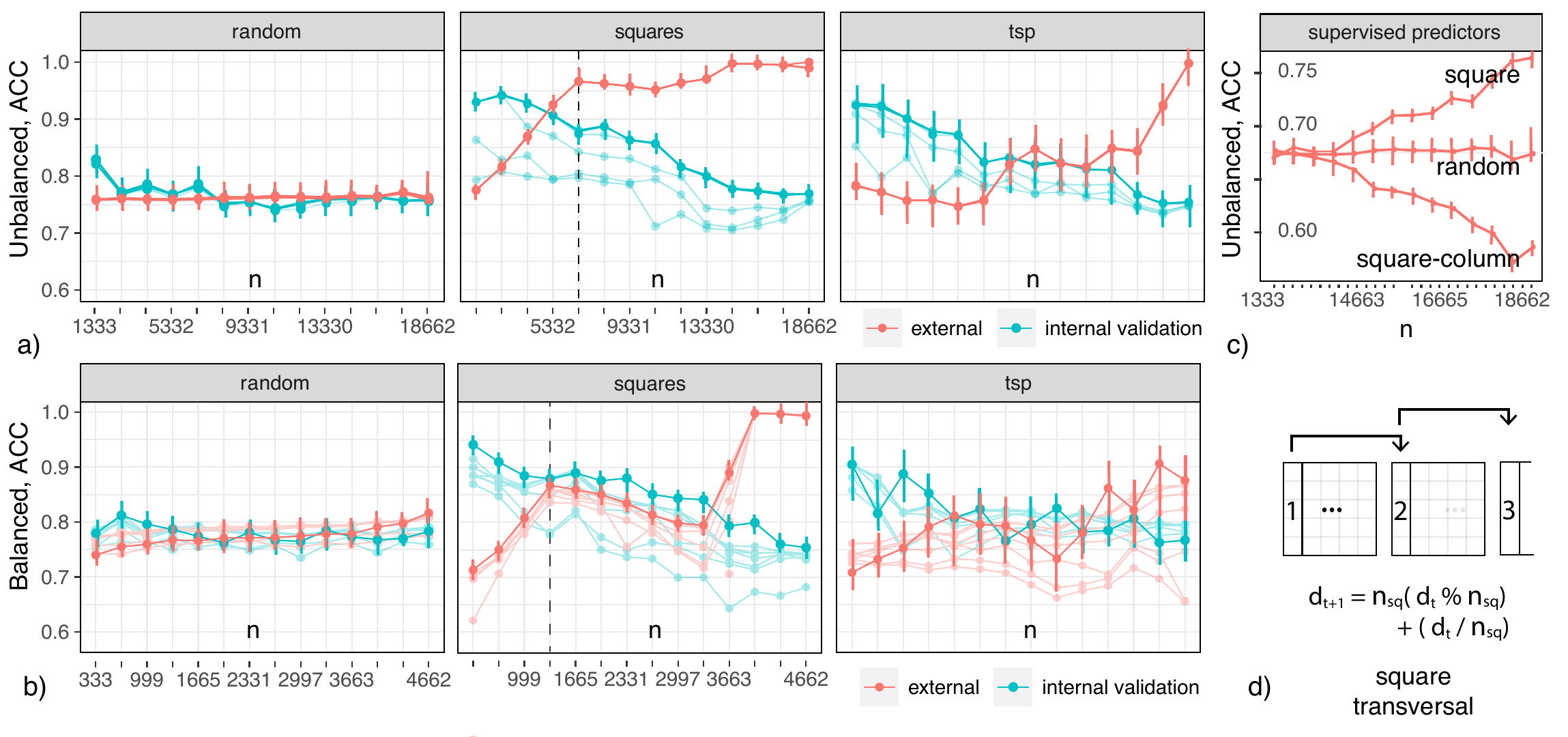}\\
\caption{supervised ACC vs. $n$ for the \textbf{(a)} \emph{unbalanced} and \textbf{(b)} \emph{balanced} cases; \textbf{(c)} ACC under different transversals; \textbf{(d)} a square transversal.}\label{fig-uncorr}
\end{figure}

An understanding of the relationship among samples and estimators' accuracy and EV is necessary to devise accurate and maximally general subsamples. To that end, we consider ACC (percentage of correctly classified cases in sample validation sections) as we increase $n$. We divide samples in two random sections: \emph{internal} and \emph{external}. The internal section is divided in typical training and validation subsections. We report accuracy of cross validation in the internal section (green) and external (red), with $4$ folds across cases. When ordered arbitrarily, or randomly, the internal-external division is inconsequential (as long as individual sections contain enough samples to cross algorithm-specific overfitting thresholds).  We will define, however, alternative sample transversal orders, and consider how they impact learning performance. Starting with a small $n$, we train several algorithms with increasing $n$, and under different orders. With each unitary increase of $n$, an observation in the external section is transferred to the internal, and the problem of model and effect generalization becomes easier. Accordingly, a sample that can generalize its model and effects to external populations the earliest can be said to have higher EV. We use: three pre-specified XGBoost GBM (Gradient Boosting Machine) models, a grid of GLMs (Generalized Linear Model), a Random Forest (DRF), LASSO and Ridge regressions, five pre-specified GBMs, a near-default Deep Neural Net, an Extremely Randomized Forest (XRT), a random grid of XGBoost GBMs, a random grid of GBMs, and a random grid of Deep Neural Nets, as well as Stacked Ensembles with all previous models. Searched and optimized parameters in each case are listed in \textit{Appendix.\ref{sect-app-methods}}.


We start with the case of increasing EV (i.e., ACC in the external section). We expect a relationship between the number of squares in samples and their EV. According to \textit{Proposition (\ref{prop-varordinal})} and Eq.(\ref{eq-gap-linear}), the square transversal order with fastest increase in ACC corresponds to a leftmost-column first, $X_{(1)}$, transversal. This transversal is illustrated in Fig.\ref{fig-uncorr}(d). We will consider other transversals, and their effects in respect to EV and CF, below. Performance of algorithms in samples with increasing sizes, under random and square transversals, are shown in Fig.\ref{fig-uncorr}(a,b) for, respectively, the \emph{unbalanced} and \emph{balanced} cases. ACC Curves for 1000 simulation runs are shown for {all} algorithms (best parameter set), but bold curves mark the leader (best algorithm and parameter-set combination). Error bars are shown only for the leader. For the random transversal case, ACC (both external and internal) remain at constant levels. Since sampling is random, algorithms minimize in this case the empirical, expected error in sample populations - which can reflect both population frequencies and sample selection biases. For the square case, there is a steep (linear) increase in EV. This indicates that maximizing the number of squares in data can lead to increases in EV.  In these cases, there is little difference in performance among different algorithms, with low variance across runs (bars), suggesting analytic limits and explanations. Recommended sample sizes for the many-squares asymptotic case, Eq.(\ref{eq-n}), are marked with a dotted line across figures (mean over runs). 


\begin{figure}
\centering
\includegraphics[width=0.8\linewidth]{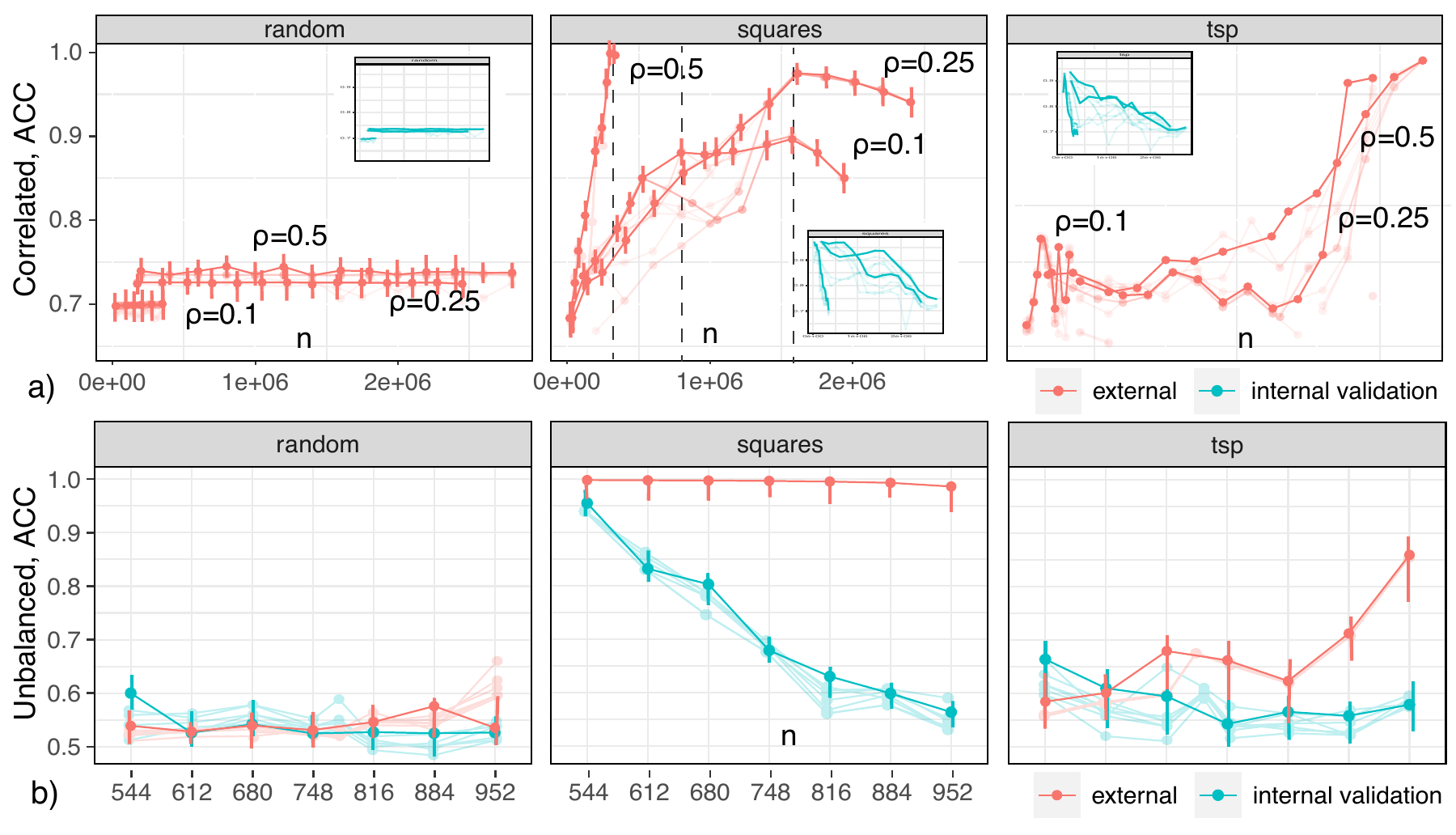}\\
\caption{ supervised ACC vs. $n$ for the \textbf{(a)} \emph{correlated}, $\rho=\{0.1,0.25,0.5\}$, and \textbf{(b)} \emph{counterfactual} prediction cases.}\label{fig-corr}
\end{figure}

The figure also shows an alternative 'Travel-Salesman' order (rightmost). It is obtained by solving a TSP\footnote{Metric TSP ~\citep{Cormen:2001aa}(pg. 1029), in the worst case it generates solutions that are twice as long as the optimal tour (TSP calculation repeated in every simulation run). }: sample units are cities, and their factor difference counts are distances. This order shares an important characteristic with square transversals: differences between units are kept small in this sampling order. The orders differ, however, in an important way: differences in the TSP are arbitrary and non-cyclic (following empirical sample frequencies). In this case, EV is steeply reducing, in contrast to square enumerations, Fig.\ref{fig-uncorr}(a,b). This is not due to algorithms becoming sensitive to sample noise - as typical in overfitting. It portrays the expected behavior from algorithms when supplied with increasingly varying inputs, generating models from specialized to general. The algorithmic stack includes regularized solutions (LASSO and Ridge regressions) and Targeted
Learning (\textit{Sect. \ref{eq-nonparam} \nameref{eq-nonparam}}, Super-learner training, \textit{Appendix.\ref{sect-app-methods}}). 

\subsection{Counterfactual and Correlated Outcome Prediction}

Patterns in Fig.\ref{fig-uncorr}(a,b) repeat across all considered generative models, but we observe increasingly larger ACC gains proceeding from the \emph{balanced} to the \emph{correlated} case,  Fig.\ref{fig-corr}(a)\footnote{TSP order (rightmost): only the best mean performance run is shown.}. Current supervised solutions are expected to perform well in samples whose variables are already independent and unbiased. Fig.\ref{fig-corr}(a) shows increasing EV gains under increasing maximum correlations $\rho=\{0.1,0.25,0.5\}$ (\emph{correlated}). Notice that, also as expected, ACC increases are linear in all cases (under square orders), Eq.(\ref{eq-gap-linear}), and that asymptotic sample bounds, Eq.(\ref{eq-n}), also hold across \emph{correlated} cases. Finally, this illustrates how square enumeration allows correlated data to be 'put into play' for contemporary predictive solutions.

While the concepts above can give researchers larger control over the generalizability of learning solutions, we started with the goal of studying the EV of counterfactual predictions. This case is show in Fig.\ref{fig-corr}(b). We take this to be the prediction of effect differences, $\Delta y_{ij}$, from covariate contrasts, $x_i-x_j$, for all sample unit pairs. The figure illustrates that everyday algorithms are clearly ineffective in this case. The intuition, and why the problem is rarely framed this way, was articulated above: using factor differences leads to samples with {increased pairwise overlaps and decreasing permutations}, decreasing the capability of algorithms to generalize effectively. The figure shows, however, that models can generalize effectively in samples with the stipulated combinatorial properties, Eq.(\ref{eq-sq-conditions}), as in square enumerations. This echoes the increased performance observed over correlated data, and reveals a relationship between these two problems.

\begin{figure}
\centering
\includegraphics[width=1\linewidth]{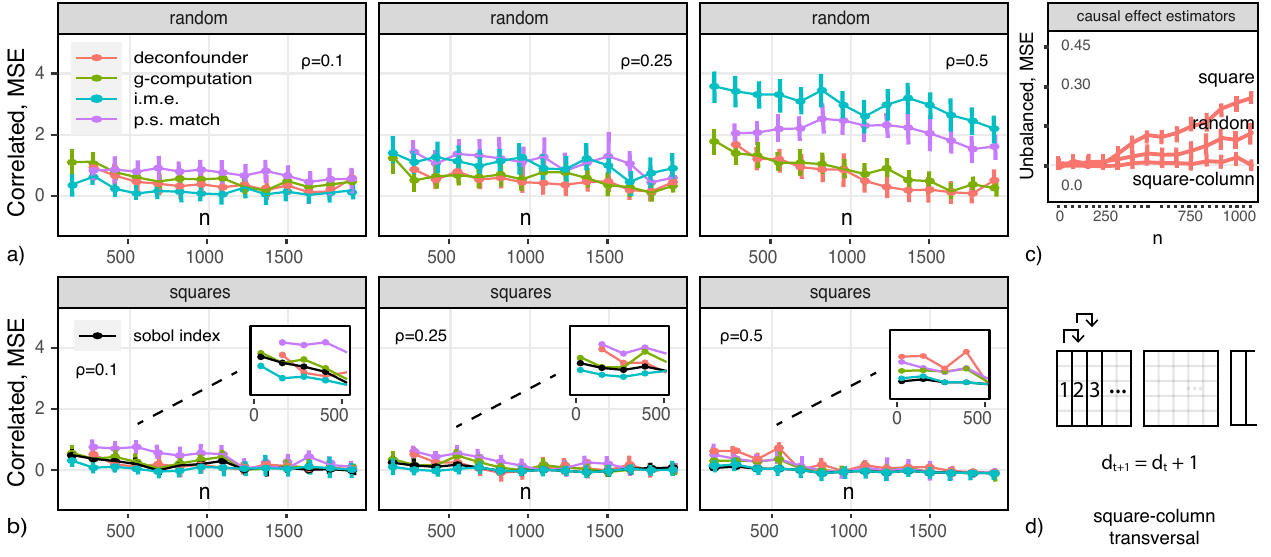}\\
\caption{causal and importance estimation  Mean-Squared Error (MSE) vs. $n$, $\rho=\{0.1,0.25,0.5\}$, under \textbf{(a)} random order, \textbf{(b)} square-column order; \textbf{(c)} MSE under distinct transversals; \textbf{(d)} a square-column transversal.}\label{fig-causal}
\end{figure}

\subsection{Causal Effect Estimation and Black-box Explanations}

Supervised solutions are highly tuned to generalizability. We now consider implications to causal techniques, which, instead, emphasize biasedness and estimation under non-i.i.d. conditions.  Fig.\ref{fig-uncorr}(d) and Fig.\ref{fig-causal}(d) illustrate two distinct ways to transverse enumerated squares. Each individual square column corresponds to sets of effect observations with a given ordinal rank, and, thus, similar confounding levels and variance, \textit{Proposition (\ref{prop-varordinal})}. Distinct square transversals reproduce EV increasing (\textit{Definition (\ref{def-ev})}) and CF non-increasing (\textit{Definition (\ref{def-cf})}) sequences. Imagine we place squares side-by-side, and that each square cell contains a sample unit. We then transverse squares from left-to-right and top-to-bottom. The right-ward sequence of columns is, however, generated in one of two ways: across squares (a \emph{square} transversal, like in the previous sections) and within squares (a \emph{square-column} transversal). Using input from these two transversals should make estimators behave asymmetrically in respect to EV and CF.  A square transversal corresponds to an order with increasing EV and minimal CF increase. A square-column transversal corresponds to an order with decreasing CF and minimal EV increase. The enumeration production rule for a column $d_t$, at time $t$, for each case is also shown in the figures, $d_1=1$. Fig.\ref{fig-uncorr}(c) shows how ACC changes for supervised systems under random, square, and square-column transversals. Fig.\ref{fig-causal}(c) shows how ACC changes, instead, for contemporary causal effect estimators and explaining systems (discussed below). Accuracy is, in this case, sum of squared differences between estimated and ground truth single-variable effects (for all variables). For a set of enumerated squares, for references $x_0, x_1,..., x_t$ at time $t$, accuracy is calculated for populations $x_0,x_1,...,x_t$. Lines correspond to mean ACC (and errors) of the best performing algorithm and parameter set (\emph{leader}) across 1000 runs. Both transversals (square and square-column) supply systems with equally-represented populations but their performance is symmetric in respect to ACC. The random sample ordering strikes a balance between these extremes. 

\begin{figure}
\centering
\includegraphics[width=1\linewidth]{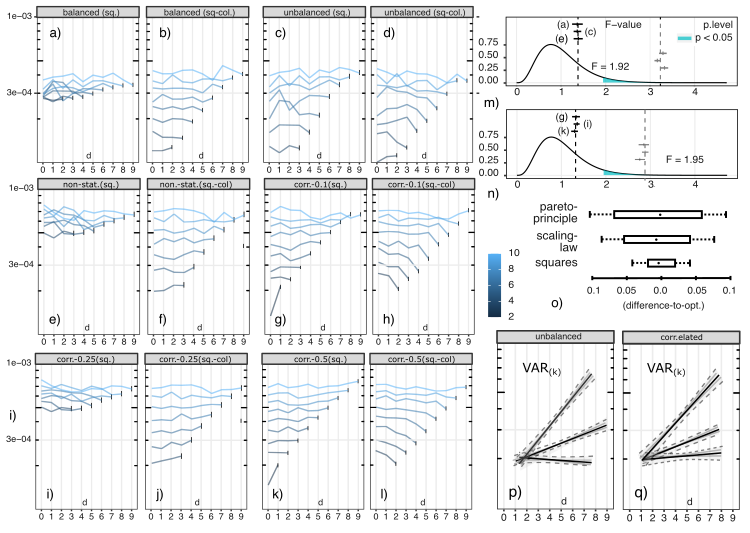}\\
\caption{pairwise empirical errors, $(\Delta_{ij}y-\Delta_{uv}y)^2$, (y-axis) with increasing square sizes $m'$ (color) and columns $d$ (x-axis), $d\leq m$, \textbf{(a-d)} \emph{balanced} and \emph{unbalanced} cases, \textbf{(e-f)} \emph{non-stationary} case,  \textbf{(g-l)} \emph{correlated} cases (log-scale);  \textbf{(m)} F-tests for \emph{balanced} and \emph{unbalanced} cases, and \textbf{(n)} for other cases (grey show results for subsamples passing tests in Eq.(\ref{eq-sq-conditions}), \emph{omitted-correlated} cases; \textbf{(o)} optimal sample split sizes; linear regressions, and confidence intervals (dashed ribbon), of effect error estimates for \textbf{(p)} \emph{omitted} and \textbf{(q)} \emph{omitted-correlated} cases.}\label{fig-error}
\end{figure}

Fig.\ref{fig-causal}(a) shows performance of 3 widely used causal effect estimators and the IME explainer ~\citep{10.5555/3295222.3295230,lundberg2019consistent} with random (top) and square-column (bottom) orders for three levels of sample correlation (\emph{correlated}, square-column transversal)\footnote{with $n<100$, omitted datapoints are due to regression non-convergence.}. The first estimator is the recent Deconfounder ~\citep{Wang:2019aa}(Sect. 3.1, linear Bayesian factor model fit with Variational Bayes,  logistic outcomes, and Normal priors). The second is  g-formula ~\citep{Chatton:2020aa}, an efficient solution popular in Epidemiology (\textit{Sect. \ref{eq-nonparam} \nameref{eq-nonparam}}). IME uses the Shapley value calculated from the output of all supervised methods in the last section. A propensity score matching estimator ~\citep{ROSENBAUM:1983aa} is included for its popularity and baseline significance in counterfactual solutions.  These algorithms (causal and explainer) make different assumptions. The first focuses on biasedness, the second combines (biased and often heuristic) predictive regressions for model selection. As expected, causal effect estimators perform well across \emph{correlated} cases, compared to the explainer, which makes i.i.d. assumptions, Fig.\ref{fig-causal}(a). Effect estimators loose accuracy with increasing correlation, however, as increasingly less data is uncorrelated in samples and used by these methods. 

The approach here also establishes a connection to variance-based sensitivity analysis ~\citep{Benesse:2024ug}. Sensitivity analysis  decompose systems' error to estimate individual factor contributions. The Sobol index ~\citep{Sobol:2001aa} is defined from the conditional variance of a factor, $\nicefrac{\text{VAR} \Expect [ y | x_i] }{\text{VAR}(y)}= \nicefrac{\text{VAR}(\beta)}{\text{VAR}(y)}$, where $\beta$ are effect estimates (often from Analysis of Variance, ANOVA). Most explaining and sensitivity analysis approaches start with a functional for outcomes, $y = f(a,b,c,...)$ where input factors $a,b, c, ...$ are i.i.d.  Black-box explainers subject boxes (algorithms) to a large number of variations, while  sensitivity analysis decomposes their errors. The goal for both is to indicate the importance of individual input variables, especially in respect to sample predictive performance. The Shapley value comes from coalitional game theory and was devised axiomatically. Beyond its strategic origins, Shapley's axioms are important requirements to decompose changes in any system ~\citep{Boer:2020aa}.  As it is often used with random sampling, it is closely related to bootstrap and permutation methods for black-box regressions. 


Squares were used here as means to {represent sets of effect observations for a unit $x_0$, their permutations in samples}, and, finally, to define non-parametric statistics for sample factors' effects.  Our initial definition of effects, Eq.(\ref{eq-eff-def}), led to consequent definitions and interpretations for the EV, \textit{Definition (\ref{def-ev})}, and CF, \textit{Definition (\ref{def-cf})}, of counterfactual observations and samples. The unit of analysis here is $\Delta y$, an outcome difference observation, and not the outcomes of regressions. Permutations are not random sample intake orders, but 'observed' permutations in samples that fulfilled a set of combinatorial and gap conditions, and offer additional guarantees over effect estimates. Eq.(\ref{eq-eff-def}) thus uses the complete {set of $m$-way observed differences} in a sample to permute {observed counterfactual outcomes} and derive effect estimates. The full set of effect background variation, $\Pi_n(Z / \{a\})$, is the full set of EV-increasing sequences in samples (or their 'external variation') for individual sample populations.


 Two fundamental problems with both the previous explaining and sensitivity analysis approaches are the need for parametric forms for outcomes, and the assumption of i.i.d. variables ~\citep{Aas:2021tx}. The latter is also a problem for standard non-parametric tests and generic permutation-based tests ~\citep{Hoeffding:1948aa,Yamato:1986aa}. Fig.\ref{fig-causal}(a) shows that correlations introduce considerable biases into these methods' estimates of variable importance. Fig.\ref{fig-causal}(b) shows, however, that both methods perform well under square enumerations. Performance increases in both fronts (effect estimation and importance attribution) in this case. In fact, in small samples, the latter dominate recent methods designed specifically for causality and correlated data (upper-right panels). This illustrates that, when samples are limited and effect estimates not externally valid, solutions aimed at generalization (given their inputs are not biased) can also offer gains in causal effect estimation tasks. The Shapley (teal curve) and Sobol indices (black) largely coincide under these conditions, as \textbf{variance becomes an unbiased indicator of importance and EV}, Eq.(\ref{eq-ev-def}). This illustrates that simple estimators (which do not involve over 20 supervised methods, repeated many times for explanations) can also become effective under square transversals.  Notice that both the Deconfounder and g-formula are highly-tuned to DGP assumptions, as both rely on logistic regressions. This exemplifies, in turn, the attraction of non-parametric insights, and that, even when models are correctly specified, there is room for gains with the proposed framework. 
 
 Altogether, these results suggest that sample size increase is not the only factor that can lead to increases in the generalizability of algorithms. The number of EV-increasing sequences in the sample play a role. At the same time, the number of CF non-increasing sequences play a role for causal effect estimation (or other tasks related to the estimation of individual factor effect). This observation is in direct contrast with, and suggests significant penalties for, the i.i.d. assumptions of black-box and bootstrap sampling approaches. These methods can observe gains, and control desired statistical tradeoffs, by sampling squares, instead of individual observations.

\subsection{Square Gap and Column Statistics}

 The panels in Fig.\ref{fig-error} (a-l) illustrate the proposed empirical error decomposition for counterfactual effect observations, Eq.(\ref{eq-gap-linear}), in enumerated sample squares. These diagrams depict the mean empirical error among pairs, $(\Delta_{ij}y-\Delta_{uv}y)^2$, (y-axis) in their $d$-th column (x-axis). These are \emph{observed} errors among sample units' outcomes (in the same square and column), and not effect estimates.  The top row, Fig.\ref{fig-error}(a-d), shows errors in \textit{balanced} and \textit{unbalanced} simulations (both transversals). Larger $d\leq m'$ lead to pairs exposed to increasing unobserved confounders, and larger variance in effects, for a fixed factor and square reference unit $x_0$. Square and square-column transversals are shown.  As expected, Eq.(\ref{eq-gap-linear}), increasing square sizes then lead to error decreases among its counterfactual observations at regular and linear rates (lines with different colors). These errors, $\textrm{VAR}_{(k)}$, are depicted by the approximately constant separation between lines. They offer effect estimates under increasing factor variation and EV, Eq.(\ref{eq-ev-def}). 
 
 


 F-tests are routinely used to compare variance across sample groups and can be used to demonstrate the proposed (ex post) sample stratification strategy. F-tests follow straight-forwardly from the previous error decomposition, Eq.(\ref{eq-gap-linear}) (\textit{Appendix.\ref{sect-bayesian}}).  Fig.\ref{fig-error}(m) shows, in particular, results of F-tests for effect variance in units from the same populations in cases (a-d) (non-stratified data). It shows that F-tests lead to false-negative identification of effects with same ground-truth from effect variances (outside the statistic $5\%$ significance level) across all cases (horizontal bars are whisker box-plot ranges). The figure also shows the result of tests for subsamples that pass the tests in Eq.(\ref{eq-sq-conditions}) (gray), which lead, instead, to true-posititive outcomes in the same simulated data. Fig.\ref{fig-error}(n) reproduces these results for the non-stationary case, where the correction is more significant.  These results are possible due to the limit on effect background association illustrated in Fig.\ref{fig-intro2}(b). Fig.\ref{fig-error}(p-q) shows Bayesian ANOVA regressions, and their confidence intervals (dashed ribbons), for  $\textrm{VAR}_{(k)}$ across 1000 runs of unbalanced and correlated cases. Different from the previous, simulations had $3$ \emph{fixed} levels of unobserved confounders ($q=\{0,10,20\}$), the ground-truth variance in effect observations according to Eq.(\ref{eq-gap-linear}) is shown in black. ANOVA regressions over the full sample (not passing the proposed combinatorial and gap tests) are non-convergent.  This illustrates that, combined with the previous tests, these simple regressions and statistics can help researchers better understand the completeness, and possible generalizability, of their samples and effect estimates. 



\subsection{The 80/20 Sample Split}

The most universally used rule-of-thumb in ML practice is the 80/20 ratio when splitting samples into train and test subsamples. Although it has remained at the theoretical margins of ML research, largely as application to the problem of overtraining in neural networks, a natural question is: why this proportion? There are two common answers. The first is heuristic, as a consequence of the Pareto Principle. The principle states that '80\% of effects come from 20\% of causes' ~\citep{Chen:1994aa}. Causes are used here vaguely, but the principle explains many natural and artificial phenomena. The second answer is more precise, formulated as a scaling-law ~\citep{Guyon97ascaling,bahri2021explaining}, generalizing ~\citep{10.5555/2998828.2998853,10.5555/2998828.2998854}. They find that 'the fraction of patterns reserved for the validation set should be inversely proportional to the square root of the number of free adjustable parameters'. In essence, the optimal split is therefore determined by the number of unique factors in a sample, and not its gross number of observations.  Eq.(\ref{eq-n})(multiple-squares) derived simple bounds, $\tilde{n}$, for sample sizes. The pareto principle and ratio scaling-law offer similar recommendations.  Fig.\ref{fig-error}(o) shows mean difference between recommended sizes, $\tilde{n}$, and the optimal split, $n_{opt}$, across all previous simulation cases. In each case, a sample with $n'>100$ observations is divided into 100 random and increasing subsamples. Each subsample adds $n'/100$ new observations to the previous and is used as training for the previous supervised methods. The optimal split $n_{opt}$ is taken as the subsample size with an inflection point in the best performance (ACC) across all methods (all samples had ACC peaks, if non-unique the smallest size was taken as optimal). Error is thus $(\tilde{n}-n_{opt})/n'$. Although an asymptotic approximation, relating to the asymptotic number of permutations in unbalanced samples, Eq.(\ref{eq-n}) describes well trade-offs in relation to the EV of training samples. 


\begin{figure}
\centering
\includegraphics[width=0.8\linewidth]{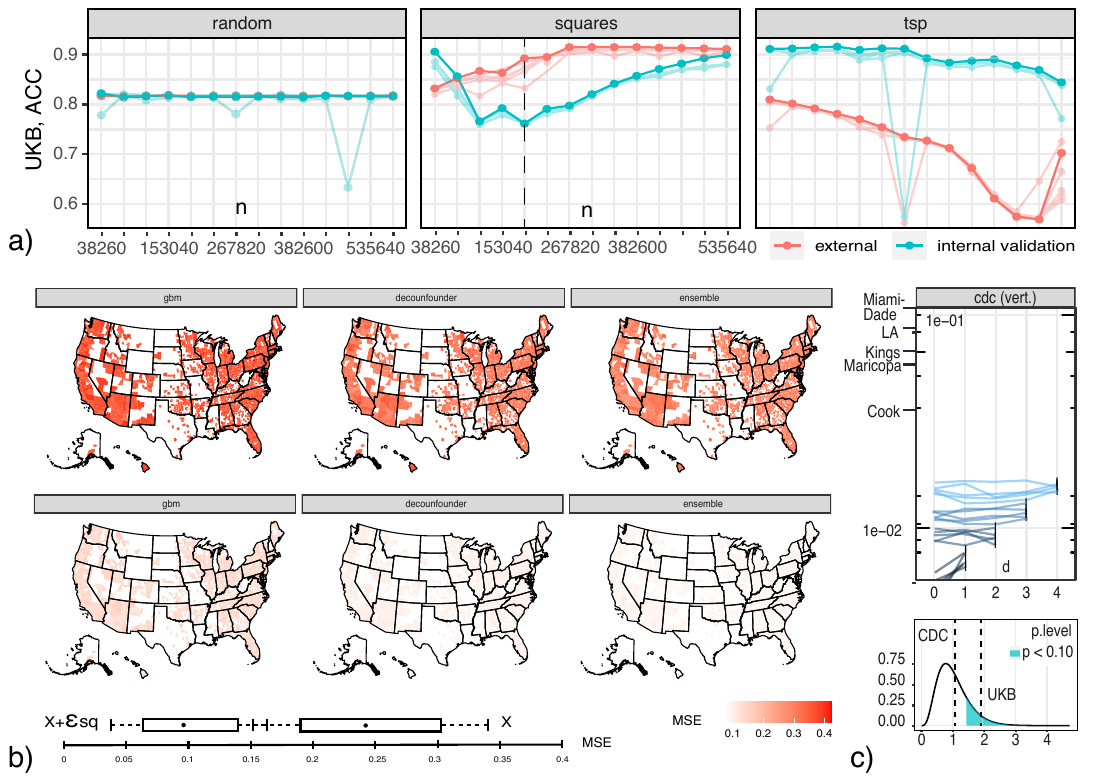}\\
\caption{\textbf{(a)} ACC vs. $n$ COVID19 infection out-of-sample prediction in the UK Biobank (UKB); \textbf{(b)} Leave-one-out prediction MSE in the Center for Disease Control dataset (CDC)(upper maps), with added $\varepsilon_{sq}$ error component (lower maps); \textbf{(c)} infection error decomposition for the most-impacted US counties (log-scale).}\label{fig-covid}
\end{figure}

\subsection{Real-World Example}\label{sect-real}
 The COVID19 pandemic continues to threaten the lives and livelihoods of billions around the world. As of the break of the pandemic, there was a rush and expectations for Machine Learning (ML) solutions to help inform policy and individuals' decisions ~\citep{10.3389/fpubh.2020.587937,Jin:2021aa,Verity:2020aa}.   In practice, few ML approaches were truly useful, with SIR semi-deterministic models ~\citep{Cramer:2021aa}, or their specializations, often favored. This is, in part, due to the few guarantees current supervised solutions can offer in respect to sample biases and heterogeneity, especially when samples are limited (which coincide with when predictions matter the most). SIR models offer predictions at high aggregate levels - most often country and city levels - taking all individuals therein to be the same. Some sources of heterogeneity were identified, age being the most obvious. We discuss in particular, how to make predictions with moderate and severe number of unobservabed factors and small $n$. We consider two data sources. The first is the UK Biobank ~\citep{Bycroft:2018aa} (UKB): a dataset with $\sim$500K UK citizens, 100K infected, 5K variables. It includes a wide variety of variables - not only from individuals' electronic medical records, but also sociodemographic, economic, living, behavior and psychological annotations. The attraction is its completeness, high-dimensionality, and level (individual). We also consider a data dump from the American Center for Disease Control (CDC) ~\citep{cdc-dataset} with $\sim$24M cases but limited variables (age, geographic, race, ethnicity, and sex). The interest is as platform to discuss the severe unobserved factors case. Binary outcome is COVID19 infection in both cases. Both datasets include information up to April 1, 2021.
 
Fig.\ref{fig-covid}(a) shows performance of the previous supervised solutions in the UK Biobank ~\citep{Bycroft:2018aa}. These achieve out-of-sample ACC of $\sim$80\% with random sampling. We see, under square ordering,  a pattern in EV similar to those in the simulated data. Taking a fourth of the sample leads to 10\% increase in ACC over the unseen data. As before, the TSP ordering leads to a sharp decrease in EV in the first-half.  At a fourth of the sample, we have larger confidence over the amount of population predictive coverage the trained models offer. The 10 variables with highest EV, according to Eq.(\ref{eq-ev-def}), and their biobank codes (parenthesis) were: \textbf{overcrowded-household} (26432), [\textbf{pop-density} (20118), traffic-intensity (24011), time-to-services (26433)], [\textbf{health-rank} (2178), smoking (20116), expiratory-flow (3064), age (24)], \textbf{job-physical} (816), \textbf{risk-taking} (2040). The first is a deprivation index marking individuals in overcrowded households. This is in line with research showing that infection for families happened in a logical-OR fashion (i.e., members multiply their individual exposures to the disease) ~\citep{Rader:2020aa,10.1001/jama.2020.11370}.  We demonstrated that the presence of many enumerable permutations in samples allow correlated factors, and confounders in particular, to be separated from causes by effect invariance, \textit{Principle (\ref{prop-sep})}. Variables in brackets have $\rho>0.1$, and are listed in order of effect invariance. There are many squares of size $10$, which indicates that, despite correlations, there are enough observations to also parse out correlated in-sample factors' effects (that pass out-of-sample gap tests). Population density is favored over neighborhood traffic intensity, and general health over age, for example. Further description of these variables can be found on ~\citep{ukb:datashowcase} from their codes.

Most data used for prediction after the pandemic onset were, however, simpler, with only case counts and a few demographic variables. The COVID19 Case Surveillance database ~\citep{cdc-dataset} includes patient-level data reported by US states to the CDC. We use the subsample with simultaneously non-missing age, sex, race, ethnicity and location (county-level). This leads to 10 binary variables and $\sim$11M cases. We additionally generated a set of baseline controls from the corresponding variables in the 2018 American census (county level). Fig.\ref{fig-covid}(c) shows $F$-test estimates for the UKB and CDC samples. Compared to the CDC dataset, the UKB has a comprehensive set of variables and $F$ under $10\%$ significance level. The estimation indicates that the CDC sample has very large numbers of omitted variables. Fig.\ref{fig-covid}(c) also repeats the previous omitted variable plots for the CDC dataset and its 5 counties with highest case counts (top). As expected, $\textrm{VAR}_{(k)}$ ANOVA estimates, Eq.(\ref{eq-gap-linear}), vary significantly across locations (upper-left, with county names) but these have little impact over observed variable pairwise errors (lower lines), as they affect observed square subpopulations uniformly and can be proxied-out, Eq.(\ref{eq-gap-linear}). The perspective suggests that, regardless of the CDC sample's severe shortcomings, errors $\textrm{VAR}_{(k)}$ can be useful by indicating degree of 'noise' and EV, Eq.(\ref{eq-ev-def}) in effect observations. Fig.\ref{fig-covid}(b)(top maps) shows Mean-Squared Error (MSE) of a Leave-One-Out task for the 1489 American counties in the CDC sample, using 3 of the best performers in the previous tasks. Case information from all \emph{other} counties is used to predict a \emph{given} location's COVID19 incidence (addressing the question of which other locations' infection information could be used to derive optimal estimates for a specific location). MSE is the mean squared difference between the predicted and (held-out) local incidence. Fig.\ref{fig-covid}(b)(lower maps) repeats the task adding $\textrm{VAR}_{(k)}$ to the inputs of these state-of-the-art estimators, which leads to significant gains across locations and algorithms. The lower box-and-whisker plot summarizes results  for the Gradient Boosting Machine ~\citep{Friedman:2000aa} (left-most maps). This illustrates that, even in clearly confounded samples, where it would be unadvised to assign relative importance to any factor, EV estimates can still carry useful information, as error estimates, for subsamples and locations. Other real-world examples are discussed in ~\citep{ribeiro-growth,ribeiro-periodic}. 

\section{Discussion and Conclusion}

The central methodological challenge in the Sciences, Policy-making  and Design remains the evaluation of counterfactual statements (Did this treatment caused the result of interest? Did this policy?)  The counterfactual definition of effects ~\citep{Rubin:2005aa,Morgan:2007aa} formulate sample properties necessary for effects to be free of selection biases, Eq.(\ref{eq-rubin}). It is based on the fixation of effect backgrounds in samples. The strategy in this work has been, in contrast, to focus on conditions where the amount of unobserved variation is guaranteed to be large (sampling is EV-increasing) and shared among factors (CF non-increasing). This alternative definition, as effects that remain invariant in the face of large external variation, can lead, instead, to generalizeable effect estimates. Under these conditions, causes can also be separated from confounders using effect variance estimates, \textit{Principle (\ref{prop-sep})}. These considerations led to the notion of effect background randomization, and sample conditions under which effects become sufficiently randomized.  The perspective also allowed us to reconsider problems that have proven difficult for the standard counterfactual formulation, such as their sample size requirements and unobserved variable biases.

The model allowed us to relate sample incompleteness to loss of estimation performance. We formulated sample properties necessary for effect External Validity, Eq.(\ref{eq-eff-def}), and their required sample sizes. They were used to define  EV and variance {statistics for individual effect} observations, as well as for sets of effect observations (samples).   Several concepts, only loosely justified in mainstream ML, such the as the hyperbolic functions ~\citep{ribeiro-growth,LeCun:aa} of neural networks, transformation group invariance ~\citep{10.5555/3322706.3322731,10.5555/3045390.3045705}, and 80-20 train-test sample split appears naturally in this purely combinatorial perspective. Cyclic permutations, and treatment-background 'convolutions', Eq.(\ref{eq-convolution}), suggest a bridge between combinatorial and continuous multi-scale or frequency-based representations. The latter consists of one of the theoretical underpinning of neural networks and several other methods.  This suggests a possible direction for the long-sought connection between causality and mainstream ML techniques ~\citep{Scholkopf:2021aa}.  We focused on confounding, external validity, and selection bias. Other concepts from the causal inference literature could be similarly implemented.  

An advantage of the proposed approach is thus bringing causal and predictive concepts closer. We demonstrated generalizability bounds for popular supervised predictors and black-box explainers \citep{10.1613/jair.1.12228}.  Supervised prediction, black-box explanation and effect estimation approaches should state, side-by-side with their estimates, the conditions on which they are expected to hold (sample sizes, correlations, completeness, etc.). This is unknown for most current causal effect estimation solutions, IME explainers, and most out-of-sample prediction solutions. Until these are stated, it is difficult to, in practice, trust their outputs. The approach allows us to model effects in both linear and non-linear data without the, rarely know and difficult to infer, parametric models for outcomes or independence among factors. We discussed out-of-sample, counterfactual, correlated and omitted variables prediction, as well as causal effect estimation, in simulations and an important real-world example. Correlated and counterfactual  prediction observed, in particular, significant accuracy gains. A central result demonstrated here is that \textbf{the EV of interventions are, to some extent, predictable from combinatorial properties of the populations they act upon}. To a broad audience, Machine Learning methods are inherently limited due to their lack of generalizability and selection bias guarantees, when compared to, for example, experimental hypothesis testing. We believe the work could help build further connections between predictive and causal techniques.


\bibliography{bib.bib}

\begin{thebibliography}{68}
\providecommand{\natexlab}[1]{#1}
\providecommand{\url}[1]{\texttt{#1}}
\expandafter\ifx\csname urlstyle\endcsname\relax
  \providecommand{\doi}[1]{doi: #1}\else
  \providecommand{\doi}{doi: \begingroup \urlstyle{rm}\Url}\fi

\bibitem[Aas et~al.(2021)Aas, Jullum, and L{\o}land]{Aas:2021tx}
K.~Aas, M.~Jullum, and A.~L{\o}land.
\newblock Explaining individual predictions when features are dependent: More
  accurate approximations to shapley values.
\newblock \emph{Artificial Intelligence}, 298:\penalty0 103502, 2021.
\newblock \doi{https://doi.org/10.1016/j.artint.2021.103502}.
\newblock URL
  \url{https://www.sciencedirect.com/science/article/pii/S0004370221000539}.

\bibitem[Abadie(2021)]{10.1257/jel.20191450}
A.~Abadie.
\newblock Using synthetic controls: Feasibility, data requirements, and
  methodological aspects.
\newblock \emph{Journal of Economic Literature}, 59\penalty0 (2):\penalty0
  391--425, June 2021.
\newblock \doi{10.1257/jel.20191450}.
\newblock URL \url{https://www.aeaweb.org/articles?id=10.1257/jel.20191450}.

\bibitem[Abadie and Imbens(2006)]{Abadie:2006aa}
A.~Abadie and G.~W. Imbens.
\newblock Large sample properties of matching estimators for average treatment
  effects.
\newblock \emph{Econometrica}, 74\penalty0 (1):\penalty0 235--267, 2006.
\newblock \doi{10.1111/j.1468-0262.2006.00655.x}.

\bibitem[Amari et~al.(1995)Amari, Murata, M\"{u}ller, Finke, and
  Yang]{10.5555/2998828.2998853}
S.~Amari, N.~Murata, K.-R. M\"{u}ller, M.~Finke, and H.~Yang.
\newblock Statistical theory of overtraining: Is cross-validation
  asymptotically effective?
\newblock In \emph{Proceedings of the 8th International Conference on Neural
  Information Processing Systems}, NIPS'95, pages 176--182, Cambridge, MA, USA,
  1995. MIT Press.

\bibitem[Bahri et~al.(2021)Bahri, Dyer, Kaplan, Lee, and
  Sharma]{bahri2021explaining}
Y.~Bahri, E.~Dyer, J.~Kaplan, J.~Lee, and U.~Sharma.
\newblock Explaining neural scaling laws, 2021.

\bibitem[Bayer and Diaconis(1992)]{bayer1992trailing}
D.~Bayer and P.~Diaconis.
\newblock Trailing the dovetail shuffle to its lair.
\newblock \emph{The Annals of Applied Probability}, 2\penalty0 (2):\penalty0
  294--313, 1992.

\bibitem[B{\'e}nesse et~al.(2024)B{\'e}nesse, Gamboa, Loubes, and
  Boissin]{Benesse:2024ug}
C.~B{\'e}nesse, F.~Gamboa, J.-M. Loubes, and T.~Boissin.
\newblock Fairness seen as global sensitivity analysis.
\newblock \emph{Machine Learning}, 113\penalty0 (5):\penalty0 3205--3232, 2024.
\newblock \doi{10.1007/s10994-022-06202-y}.
\newblock URL \url{https://doi.org/10.1007/s10994-022-06202-y}.

\bibitem[Bietti and Mairal(2019)]{10.5555/3322706.3322731}
A.~Bietti and J.~Mairal.
\newblock Group invariance, stability to deformations, and complexity of deep
  convolutional representations.
\newblock \emph{J. Mach. Learn. Res.}, 20\penalty0 (1):\penalty0 876--924, jan
  2019.
\newblock ISSN 1532-4435.

\bibitem[Buehlmann(2020)]{Buehlmann:2020aa}
P.~Buehlmann.
\newblock Invariance, causality and robustness.
\newblock \emph{Statistical science}, 35\penalty0 (3):\penalty0 404--426, 2020.
\newblock \doi{10.1214/19-STS721}.

\bibitem[Buhler and Graham.(c2004.)]{286326}
J.~Buhler and R.~Graham.
\newblock \emph{Juggling patterns, passing, and posets}, volume Mathematical
  adventures for students and amateurs /.
\newblock Mathematical Association of America, [Washington, DC] :, c2004.

\bibitem[Burkart and Huber(2021)]{10.1613/jair.1.12228}
N.~Burkart and M.~F. Huber.
\newblock A survey on the explainability of supervised machine learning.
\newblock \emph{J. Artif. Int. Res.}, 70:\penalty0 245--317, may 2021.
\newblock ISSN 1076-9757.
\newblock \doi{10.1613/jair.1.12228}.
\newblock URL \url{https://doi.org/10.1613/jair.1.12228}.

\bibitem[Bycroft et~al.(2018)Bycroft, Freeman, Petkova, Band, Elliott, Sharp,
  Motyer, Vukcevic, Delaneau, O'Connell, Cortes, Welsh, Young, Effingham,
  McVean, Leslie, Allen, Donnelly, and Marchini]{Bycroft:2018aa}
C.~Bycroft, C.~Freeman, D.~Petkova, G.~Band, L.~T. Elliott, K.~Sharp,
  A.~Motyer, D.~Vukcevic, O.~Delaneau, J.~O'Connell, A.~Cortes, S.~Welsh,
  A.~Young, M.~Effingham, G.~McVean, S.~Leslie, N.~Allen, P.~Donnelly, and
  J.~Marchini.
\newblock The uk biobank resource with deep phenotyping and genomic data.
\newblock \emph{Nature}, 562\penalty0 (7726):\penalty0 203--209, 2018.
\newblock \doi{10.1038/s41586-018-0579-z}.
\newblock URL \url{https://doi.org/10.1038/s41586-018-0579-z}.

\bibitem[Calonico et~al.(2019)Calonico, Cattaneo, Farrell, and
  Titiunik]{10.1162/rest_a_00760}
S.~Calonico, M.~D. Cattaneo, M.~H. Farrell, and R.~Titiunik.
\newblock {Regression Discontinuity Designs Using Covariates}.
\newblock \emph{The Review of Economics and Statistics}, 101\penalty0
  (3):\penalty0 442--451, 07 2019.
\newblock ISSN 0034-6535.
\newblock \doi{10.1162/rest_a_00760}.
\newblock URL \url{https://doi.org/10.1162/rest\_a\_00760}.

\bibitem[CDC-Dataset-2021()]{cdc-dataset}
C.~S.~D. CDC-Dataset-2021.
\newblock
  https://data.cdc.gov/case-surveillance/covid-19-case-surveillance-public-use-data-with-ge/n8mc-b4w4.

\bibitem[Chatton et~al.(2020)Chatton, Le~Borgne, Leyrat, Gillaizeau, Rousseau,
  Barbin, Laplaud, Leger, Giraudeau, and Foucher]{Chatton:2020aa}
A.~Chatton, F.~Le~Borgne, C.~Leyrat, F.~Gillaizeau, C.~Rousseau, L.~Barbin,
  D.~Laplaud, M.~Leger, B.~Giraudeau, and Y.~Foucher.
\newblock G-computation, propensity score-based methods, and targeted maximum
  likelihood estimator for causal inference with different covariates sets: a
  comparative simulation study.
\newblock \emph{Nature Scientific reports}, 10\penalty0 (1):\penalty0
  9219--9219, 2020.
\newblock \doi{10.1038/s41598-020-65917-x}.

\bibitem[Chen et~al.(1994)Chen, Chong, and Tong]{Chen:1994aa}
Y.~S. Chen, P.~P. Chong, and M.~Y. Tong.
\newblock Mathematical and computer modelling of the pareto principle.
\newblock \emph{Mathematical and Computer Modelling}, 19\penalty0 (9):\penalty0
  61--80, 1994.
\newblock \doi{https://doi.org/10.1016/0895-7177(94)90041-8}.
\newblock URL
  \url{https://www.sciencedirect.com/science/article/pii/0895717794900418}.

\bibitem[Cohen and Welling(2016)]{10.5555/3045390.3045705}
T.~S. Cohen and M.~Welling.
\newblock Group equivariant convolutional networks.
\newblock In \emph{Proceedings of the 33rd International Conference on
  International Conference on Machine Learning - Volume 48}, ICML'16, pages
  2990--2999. JMLR.org, 2016.

\bibitem[Cormen(2001)]{Cormen:2001aa}
T.~H. Cormen.
\newblock \emph{Introduction to algorithms}.
\newblock MIT Press : McGraw-Hill, Cambridge, Mass.; Boston, 2001.
\newblock ISBN 0262032937.

\bibitem[Correa and Bareinboim(2020)]{NEURIPS2020_7b497aa1}
J.~Correa and E.~Bareinboim.
\newblock General transportability of soft interventions: Completeness results.
\newblock In H.~Larochelle, M.~Ranzato, R.~Hadsell, M.~Balcan, and H.~Lin,
  editors, \emph{Advances in Neural Information Processing Systems}, volume~33,
  pages 10902--10912. Curran Associates, Inc., 2020.
\newblock URL
  \url{https://proceedings.neurips.cc/paper_files/paper/2020/file/7b497aa1b2a83ec63d1777a88676b0c2-Paper.pdf}.

\bibitem[Cox(2009)]{Cox:2009tn}
D.~R. Cox.
\newblock Randomization in the design of experiments.
\newblock \emph{International Statistical Review}, 77\penalty0 (3):\penalty0
  415--429, 2024/03/03 2009.
\newblock \doi{https://doi.org/10.1111/j.1751-5823.2009.00084.x}.
\newblock URL \url{https://doi.org/10.1111/j.1751-5823.2009.00084.x}.

\bibitem[Cramer et~al.(2021)Cramer, Lopez, Niemi, George, Cegan, Dettwiller,
  England, Farthing, Hunter, Lafferty, Linkov, Mayo, Parno, Rowland, Trump,
  Wang, Gao, Gu, Kim, Wang, Walker, Slayton, Johansson, and
  Biggerstaff]{Cramer:2021aa}
E.~Y. Cramer, V.~K. Lopez, J.~Niemi, G.~E. George, J.~C. Cegan, I.~D.
  Dettwiller, W.~P. England, M.~W. Farthing, R.~H. Hunter, B.~Lafferty,
  I.~Linkov, M.~L. Mayo, M.~D. Parno, M.~A. Rowland, B.~D. Trump, L.~Wang,
  L.~Gao, Z.~Gu, M.~Kim, Y.~Wang, J.~W. Walker, R.~B. Slayton, M.~Johansson,
  and M.~Biggerstaff.
\newblock Evaluation of individual and ensemble probabilistic forecasts of
  covid-19 mortality in the us, 2021.

\bibitem[de~Boer and Rodrigues(2020)]{Boer:2020aa}
P.~de~Boer and J.~F.~D. Rodrigues.
\newblock Decomposition analysis: when to use which method?
\newblock \emph{Economic systems research}, 32\penalty0 (1):\penalty0 1--28,
  2020.
\newblock \doi{10.1080/09535314.2019.1652571}.

\bibitem[Diaconis(1996)]{Diaconis:1996vh}
P.~Diaconis.
\newblock The cutoff phenomenon in finite markov chains.
\newblock \emph{Proceedings of the National Academy of Sciences}, 93\penalty0
  (4):\penalty0 1659--1664, 2024/02/26 1996.
\newblock \doi{10.1073/pnas.93.4.1659}.
\newblock URL \url{https://doi.org/10.1073/pnas.93.4.1659}.

\bibitem[Diaconis and Fulman(2023)]{diaconis2023mathematics}
P.~Diaconis and J.~Fulman.
\newblock \emph{The Mathematics of Shuffling Cards}.
\newblock American Mathematical Society, 2023.
\newblock ISBN 9781470463038.
\newblock URL \url{https://books.google.co.uk/books?id=dB2_EAAAQBAJ}.

\bibitem[Diaconis et~al.(1983)Diaconis, Graham, and Kantor]{Diaconis:1983uo}
P.~Diaconis, R.~L. Graham, and W.~M. Kantor.
\newblock The mathematics of perfect shuffles.
\newblock \emph{Advances in Applied Mathematics}, 4\penalty0 (2):\penalty0
  175--196, 1983.
\newblock \doi{https://doi.org/10.1016/0196-8858(83)90009-X}.
\newblock URL
  \url{https://www.sciencedirect.com/science/article/pii/019688588390009X}.

\bibitem[Emeruwa et~al.(2020)Emeruwa, Ona, Shaman, Turitz, Wright,
  Gyamfi-Bannerman, and Melamed]{10.1001/jama.2020.11370}
U.~N. Emeruwa, S.~Ona, J.~L. Shaman, A.~Turitz, J.~D. Wright,
  C.~Gyamfi-Bannerman, and A.~Melamed.
\newblock {Associations Between Built Environment, Neighborhood Socioeconomic
  Status, and SARS-CoV-2 Infection Among Pregnant Women in New York City}.
\newblock \emph{JAMA}, 324\penalty0 (4):\penalty0 390--392, 07 2020.
\newblock ISSN 0098-7484.
\newblock \doi{10.1001/jama.2020.11370}.
\newblock URL \url{https://doi.org/10.1001/jama.2020.11370}.

\bibitem[F.~Ribeiro et~al.(2022)F.~Ribeiro, Neffke, and
  Hausmann]{F.-Ribeiro:2022tm}
A.~F.~Ribeiro, F.~Neffke, and R.~Hausmann.
\newblock What can the millions of random treatments in nonexperimental data
  reveal about causes?
\newblock \emph{Springer Nature Computer Science}, 3\penalty0 (6):\penalty0
  421, 2022.
\newblock \doi{10.1007/s42979-022-01319-2}.
\newblock URL \url{https://doi.org/10.1007/s42979-022-01319-2}.

\bibitem[Friedman et~al.(2000)Friedman, Hastie, and
  Tibshirani]{Friedman:2000aa}
J.~Friedman, T.~Hastie, and R.~Tibshirani.
\newblock Special invited paper. additive logistic regression: A statistical
  view of boosting.
\newblock \emph{The Annals of statistics}, 28\penalty0 (2):\penalty0 337--374,
  2000.

\bibitem[Fulman(1998)]{Fulman:1998uz}
J.~Fulman.
\newblock The combinatorics of biased riffle shuffles.
\newblock \emph{Combinatorica}, 18\penalty0 (2):\penalty0 173--184, 1998.
\newblock \doi{10.1007/PL00009814}.
\newblock URL \url{https://doi.org/10.1007/PL00009814}.

\bibitem[Grimmer et~al.(2020)Grimmer, Knox, and Stewart]{Grimmer:2020aa}
J.~Grimmer, D.~Knox, and B.~M. Stewart.
\newblock Na{$\backslash$}"ive regression requires weaker assumptions than
  factor models to adjust for multiple cause confounding.
\newblock 2020.

\bibitem[Gut(2009)]{10.5555/1717330}
A.~Gut.
\newblock \emph{An Intermediate Course in Probability}.
\newblock Springer Publishing Company, Incorporated, 2nd edition, 2009.
\newblock ISBN 1441901612.

\bibitem[Guyon(1997)]{Guyon97ascaling}
I.~Guyon.
\newblock A scaling law for the validation-set training-set size ratio.
\newblock In \emph{AT and T Bell Laboratories}, 1997.

\bibitem[Hand and Till(2001)]{Hand:2001us}
D.~J. Hand and R.~J. Till.
\newblock A simple generalisation of the area under the roc curve for multiple
  class classification problems.
\newblock \emph{Machine Learning}, 45\penalty0 (2):\penalty0 171--186, 2001.
\newblock \doi{10.1023/A:1010920819831}.
\newblock URL \url{https://doi.org/10.1023/A:1010920819831}.

\bibitem[Hanson et~al.(1983)Hanson, Seyffarth, and Weston]{Hanson:1983aa}
D.~Hanson, K.~Seyffarth, and J.~H. Weston.
\newblock Matchings, derangements, rencontres.
\newblock \emph{Mathematics Magazine}, 56\penalty0 (4):\penalty0 224--229,
  1983.
\newblock \doi{10.2307/2689812}.

\bibitem[Hoeffding(1948)]{Hoeffding:1948aa}
W.~Hoeffding.
\newblock A class of statistics with asymptotically normal distribution.
\newblock \emph{The Annals of mathematical statistics}, 19\penalty0
  (3):\penalty0 293--325, 1948.
\newblock \doi{10.1214/aoms/1177730196}.

\bibitem[Jin et~al.(2021)Jin, Agarwala, Kundu, Harvey, Zhang, Wallace, and
  Chatterjee]{Jin:2021aa}
J.~Jin, N.~Agarwala, P.~Kundu, B.~Harvey, Y.~Zhang, E.~Wallace, and
  N.~Chatterjee.
\newblock Individual and community-level risk for covid-19 mortality in the
  united states.
\newblock \emph{Nature Medicine}, 27\penalty0 (2):\penalty0 264--269, 2021.
\newblock \doi{10.1038/s41591-020-01191-8}.
\newblock URL \url{https://doi.org/10.1038/s41591-020-01191-8}.

\bibitem[Kearns(1995)]{10.5555/2998828.2998854}
M.~Kearns.
\newblock A bound on the error of cross validation using the approximation and
  estimation rates, with consequences for the training-test split.
\newblock In \emph{Proceedings of the 8th International Conference on Neural
  Information Processing Systems}, NIPS'95, pages 183--189, Cambridge, MA, USA,
  1995. MIT Press.

\bibitem[Keil et~al.(2014)Keil, Edwards, Richardson, Naimi, and
  Cole]{Keil:2014uc}
A.~P. Keil, J.~K. Edwards, D.~B. Richardson, A.~I. Naimi, and S.~R. Cole.
\newblock The parametric g-formula for time-to-event data: Intuition and a
  worked example.
\newblock \emph{Epidemiology}, 25\penalty0 (6), 2014.
\newblock URL
  \url{https://journals.lww.com/epidem/fulltext/2014/11000/the_parametric_g_formula_for_time_to_event_data_.16.aspx}.

\bibitem[Kempthorne and Doerfler(1969)]{10.1093/biomet/56.2.231}
O.~Kempthorne and T.~E. Doerfler.
\newblock {The behaviour of some significance tests under experimental
  randomization}.
\newblock \emph{Biometrika}, 56\penalty0 (2):\penalty0 231--248, 08 1969.
\newblock ISSN 0006-3444.
\newblock \doi{10.1093/biomet/56.2.231}.
\newblock URL \url{https://doi.org/10.1093/biomet/56.2.231}.

\bibitem[Knuth(1997)]{10.5555/270146}
D.~E. Knuth.
\newblock \emph{The art of computer programming, volume 2 (3rd ed.):
  seminumerical algorithms}.
\newblock Addison-Wesley Longman Publishing Co., Inc., USA, 1997.
\newblock ISBN 0201896842.

\bibitem[LeCun et~al.()LeCun, Bottou, Orr, and M{\"u}ller]{LeCun:aa}
Y.~A. LeCun, L.~Bottou, G.~B. Orr, and K.-R. M{\"u}ller.
\newblock \emph{Efficient BackProp}, pages 9--48.
\newblock Springer Berlin Heidelberg, Berlin, Heidelberg.
\newblock ISBN 0302-9743.
\newblock \doi{10.1007/978-3-642-35289-8{\_}3}.

\bibitem[Lee(1990)]{Lee:1990aa}
A.~J. Lee.
\newblock \emph{U-statistics : theory and practice}.
\newblock M. Dekker, New York, 1990.
\newblock ISBN 0824782534.

\bibitem[Li et~al.(2020)Li, Horowitz, Liu, Chew, Lan, Liu, Sha, and
  Yang]{10.3389/fpubh.2020.587937}
Y.~Li, M.~A. Horowitz, J.~Liu, A.~Chew, H.~Lan, Q.~Liu, D.~Sha, and C.~Yang.
\newblock Individual-level fatality prediction of covid-19 patients using ai
  methods.
\newblock \emph{Frontiers in Public Health}, 8:\penalty0 566, 2020.
\newblock ISSN 2296-2565.
\newblock \doi{10.3389/fpubh.2020.587937}.
\newblock URL
  \url{https://www.frontiersin.org/article/10.3389/fpubh.2020.587937}.

\bibitem[Lundberg and Lee(2017)]{10.5555/3295222.3295230}
S.~M. Lundberg and S.-I. Lee.
\newblock A unified approach to interpreting model predictions.
\newblock In \emph{Proceedings of the 31st International Conference on Neural
  Information Processing Systems}, NIPS'17, pages 4768--4777, Red Hook, NY,
  USA, 2017. Curran Associates Inc.
\newblock ISBN 9781510860964.

\bibitem[Lundberg et~al.(2019)Lundberg, Erion, and Lee]{lundberg2019consistent}
S.~M. Lundberg, G.~G. Erion, and S.-I. Lee.
\newblock Consistent individualized feature attribution for tree ensembles,
  2019.

\bibitem[Magliacane et~al.(2017)Magliacane, van Ommen, Claassen, Bongers,
  Versteeg, and Mooij]{Magliacane:2017aa}
S.~Magliacane, T.~van Ommen, T.~Claassen, S.~Bongers, P.~Versteeg, and J.~M.
  Mooij.
\newblock Domain adaptation by using causal inference to predict invariant
  conditional distributions.
\newblock 2017.

\bibitem[Montgomery(2001)]{Montgomery:2001aa}
D.~C. Montgomery.
\newblock \emph{Design and analysis of experiments}.
\newblock John Wiley, New York, 2001.
\newblock ISBN 0471316490; 9780471316497.

\bibitem[Morgan and Winship(2007)]{Morgan:2007aa}
S.~L. Morgan and C.~Winship.
\newblock \emph{Counterfactuals and Causal Inference: Methods and Principles
  for Social Research}.
\newblock Cambridge University Press, Cambridge, 2007.
\newblock ISBN 0521671930; 9780521856157; 9780521671934; 0521856159.
\newblock \doi{10.1017/CBO9780511804564}.

\bibitem[Pearl(2000)]{RefWorks:doc:5911ec76e4b0ac17f7d9f458}
J.~Pearl.
\newblock \emph{Causality : models, reasoning, and inference}.
\newblock Cambridge, U.K. ; New York, 2000.
\newblock ISBN 0521773628.
\newblock Includes bibliographical references (p. 359-373) and indexes.; ID:
  http://id.lib.harvard.edu/aleph/008372583/catalog.

\bibitem[Pearl and Bareinboim(2011)]{6137426}
J.~Pearl and E.~Bareinboim.
\newblock Transportability of causal and statistical relations: A formal
  approach.
\newblock In \emph{2011 IEEE 11th International Conference on Data Mining
  Workshops}, pages 540--547, 2011.
\newblock \doi{10.1109/ICDMW.2011.169}.

\bibitem[Pearl and Bareinboim(2014)]{Pearl:2014ug}
J.~Pearl and E.~Bareinboim.
\newblock External validity: From do-calculus to transportability across
  populations.
\newblock \emph{Statistical Science}, 29\penalty0 (4):\penalty0 579--595, 11
  2014.
\newblock \doi{10.1214/14-STS486}.
\newblock URL \url{https://doi.org/10.1214/14-STS486}.

\bibitem[Peters et~al.(2016)Peters, B{\"u}hlmann, and
  Meinshausen]{Peters:2016aa}
J.~Peters, P.~B{\"u}hlmann, and N.~Meinshausen.
\newblock Causal inference by using invariant prediction: identification and
  confidence intervals.
\newblock \emph{Journal of the Royal Statistical Society. Series B, Statistical
  methodology}, 78\penalty0 (5):\penalty0 947--1012, 2016.
\newblock \doi{10.1111/rssb.12167}.

\bibitem[Rader et~al.(2020)Rader, Scarpino, Nande, Hill, Adlam, Reiner, Pigott,
  Gutierrez, Zarebski, Shrestha, Brownstein, Castro, Dye, Tian, Pybus, and
  Kraemer]{Rader:2020aa}
B.~Rader, S.~V. Scarpino, A.~Nande, A.~L. Hill, B.~Adlam, R.~C. Reiner, D.~M.
  Pigott, B.~Gutierrez, A.~E. Zarebski, M.~Shrestha, J.~S. Brownstein, M.~C.
  Castro, C.~Dye, H.~Tian, O.~G. Pybus, and M.~U.~G. Kraemer.
\newblock Crowding and the shape of covid-19 epidemics.
\newblock \emph{Nature Medicine}, 26\penalty0 (12):\penalty0 1829--1834, 2020.
\newblock \doi{10.1038/s41591-020-1104-0}.
\newblock URL \url{https://doi.org/10.1038/s41591-020-1104-0}.

\bibitem[R{\'e}nyi(1953)]{Renyi:1953ww}
A.~R{\'e}nyi.
\newblock On the theory of order statistics.
\newblock \emph{Acta Mathematica Academiae Scientiarum Hungarica}, 4\penalty0
  (3):\penalty0 191--231, 1953.
\newblock \doi{10.1007/BF02127580}.
\newblock URL \url{https://doi.org/10.1007/BF02127580}.

\bibitem[Ribeiro(2022{\natexlab{a}})]{ribeiro-growth}
A.~F. Ribeiro.
\newblock Spatiocausal patterns of sample growth, 2022{\natexlab{a}}.
\newblock URL \url{https://arxiv.org/abs/2202.13961}.

\bibitem[Ribeiro(2022{\natexlab{b}})]{ribeiro-periodic}
A.~F. Ribeiro.
\newblock Population structure and effect generalization, 2022{\natexlab{b}}.
\newblock URL \url{https://arxiv.org/abs/2209.13560}.

\bibitem[Rosenbaum and Rubin(1983)]{ROSENBAUM:1983aa}
P.~R. Rosenbaum and D.~B. Rubin.
\newblock The central role of the propensity score in observational studies for
  causal effects.
\newblock \emph{Biometrika}, 70\penalty0 (1):\penalty0 41--55, 1983.
\newblock \doi{10.1093/biomet/70.1.41}.

\bibitem[Rubin(2005)]{Rubin:2005aa}
D.~B. Rubin.
\newblock Causal inference using potential outcomes: Design, modeling,
  decisions.
\newblock \emph{Journal of the American Statistical Association}, 100\penalty0
  (469):\penalty0 322--331, 2005.
\newblock \doi{10.1198/016214504000001880}.

\bibitem[Scholkopf et~al.(2021)Scholkopf, Locatello, Bauer, Ke, Kalchbrenner,
  Goyal, and Bengio]{Scholkopf:2021aa}
B.~Scholkopf, F.~Locatello, S.~Bauer, N.~R. Ke, N.~Kalchbrenner, A.~Goyal, and
  Y.~Bengio.
\newblock Toward causal representation learning.
\newblock \emph{Proceedings of the IEEE}, 109\penalty0 (5):\penalty0 612--634,
  2021.
\newblock \doi{10.1109/JPROC.2021.3058954}.

\bibitem[Shalit et~al.(2017)Shalit, Johansson, and Sontag]{pmlr-v70-shalit17a}
U.~Shalit, F.~D. Johansson, and D.~Sontag.
\newblock Estimating individual treatment effect: generalization bounds and
  algorithms.
\newblock In D.~Precup and Y.~W. Teh, editors, \emph{Proceedings of the 34th
  International Conference on Machine Learning}, volume~70 of \emph{Proceedings
  of Machine Learning Research}, pages 3076--3085. PMLR, 06--11 Aug 2017.
\newblock URL \url{http://proceedings.mlr.press/v70/shalit17a.html}.

\bibitem[Sobol(2001)]{Sobol:2001aa}
I.~M. Sobol.
\newblock Global sensitivity indices for nonlinear mathematical models and
  their monte carlo estimates.
\newblock \emph{Mathematics and Computers in Simulation}, 55\penalty0
  (1):\penalty0 271--280, 2001.
\newblock \doi{https://doi.org/10.1016/S0378-4754(00)00270-6}.
\newblock URL
  \url{https://www.sciencedirect.com/science/article/pii/S0378475400002706}.

\bibitem[Tibshirani et~al.(2001)Tibshirani, Friedman, and
  Hastie]{Tibshirani:2001aa}
R.~Tibshirani, J.~H. J. H.~. Friedman, and T.~Hastie.
\newblock \emph{The elements of statistical learning : data mining, inference,
  and prediction}.
\newblock Springer, New York, 2001.
\newblock ISBN 0387952845.

\bibitem[UKB-Showcase-2021()]{ukb:datashowcase}
U.~B.~D. UKB-Showcase-2021.
\newblock https://biobank.ndph.ox.ac.uk/showcase/.

\bibitem[van~der Laan and Rose(2011)]{van2011targeted}
M.~van~der Laan and S.~Rose.
\newblock \emph{Targeted Learning: Causal Inference for Observational and
  Experimental Data}.
\newblock Springer Series in Statistics. Springer New York, 2011.
\newblock ISBN 9781441997821.
\newblock URL \url{https://books.google.co.uk/books?id=RGnSX5aCAgQC}.

\bibitem[Verity et~al.(2020)Verity, Okell, and Dorigatti]{Verity:2020aa}
R.~Verity, L.~C. Okell, and Dorigatti.
\newblock Estimates of the severity of coronavirus disease 2019: a model-based
  analysis (vol 20, pg 669, 2020).
\newblock \emph{The Lancet infectious diseases}, 20\penalty0 (6):\penalty0
  E116--E116, 2020.
\newblock \doi{10.1016/S1473-3099(20)30309-1}.

\bibitem[Wang and Blei(2020)]{Wang:2019aa}
Y.~Wang and D.~M. Blei.
\newblock The blessings of multiple causes.
\newblock \emph{Journal of the American Statistical Association}, 114\penalty0
  (528):\penalty0 1574--1596, 2020.
\newblock \doi{10.1080/01621459.2019.1686987}.

\bibitem[Yamato and Maesono(1986)]{Yamato:1986aa}
H.~Yamato and Y.~Maesono.
\newblock Invariant u-statistics.
\newblock \emph{Communications in statistics. Theory and methods}, 15\penalty0
  (11):\penalty0 3253--3263, 1986.
\newblock \doi{10.1080/03610928608829307}.

\bibitem[Zhang and Zhao(2023)]{Zhang:2023to}
Y.~Zhang and Q.~Zhao.
\newblock What is a randomization test?
\newblock \emph{Journal of the American Statistical Association}, 118\penalty0
  (544):\penalty0 2928--2942, 10 2023.
\newblock \doi{10.1080/01621459.2023.2199814}.
\newblock URL \url{https://doi.org/10.1080/01621459.2023.2199814}.

\end{thebibliography}

\appendix

\section{Cause-Confounder Separation from Effect Observations }\label{sect-cf}

\section{Separation and Effect Variance}\label{sect-omitted-corr}

\section{Expected Number of Draws for All Binary Outcomes}\label{app-expectdraw}

\section{Number of Squares and Derangements}\label{app-nsq}

\section{Alternating Binomial Coefficients Sum to Zero}\label{app-altsum}

\section{Supervised Algorithms and Parameters}\label{sect-app-methods}

\section{Bayesian Hierarchical Sample Errors Estimation}\label{sect-bayesian}

\section{Sample Sizes}\label{sect-power-app}

\subsection{Multiple Squares}

\section{Enumeration}\label{sect-enumeration}

\end{document}